\documentstyle[preprint,aps,version2,epsf]{revtex}
\catcode`\@=11
\long\def\@makefntext#1{
\protect\noindent \hbox to 3.2pt {\hskip-.9pt
$^{{\ninerm\@thefnmark}}$\hfil}#1\hfill}                
\def\@makefnmark{\hbox to 0pt{$^{\@thefnmark}$\hss}}  
\def\ps@myheadings{\let\@mkboth\@gobbletwo
\def\@oddhead{\hbox{}
\rightmark\hfil\ninerm\thepage}
\def\@oddfoot{}\def\@evenhead{\ninerm\thepage\hfil
\leftmark\hbox{}}\def\@evenfoot{}
\def\sectionmark##1{}\def\subsectionmark##1{}}
\renewcommand{\thefootnote}{\fnsymbol{footnote}}
\newcounter{sectionc}\newcounter{subsectionc}\newcounter{subsubsectionc}
\renewcommand{\section}[1] {\vspace*{0.6cm}\addtocounter{sectionc}{1}
\setcounter{subsectionc}{0}\setcounter{subsubsectionc}{0}\noindent
        {\normalsize\bf\thesectionc. #1}\par\vspace*{0.4cm}} 
\renewcommand{\subsection}[1] {\vspace*{0.6cm}\addtocounter{subsectionc}{1}
        \setcounter{subsubsectionc}{0}\noindent
    {\normalsize\it\thesectionc.\thesubsectionc. #1}\par\vspace*{0.4cm}}
\renewcommand{\subsubsection}[1]
{\vspace*{0.6cm}\addtocounter{subsubsectionc}{1}
\noindent {\normalsize\rm\thesectionc.\thesubsectionc.\thesubsubsectionc.
    #1}\par\vspace*{0.4cm}}

\newcounter{appendixc}
\newcounter{subappendixc}[appendixc]
\newcounter{subsubappendixc}[subappendixc]

\renewcommand{\appendix}[1] {\vspace*{0.6cm}
        \refstepcounter{appendixc}
        \setcounter{figure}{0}
        \setcounter{table}{0}
        \setcounter{equation}{0}
        \renewcommand{\thefigure}{\Alph{appendixc}.\arabic{figure}}
        \renewcommand{\thetable}{\Alph{appendixc}.\arabic{table}}
        \renewcommand{\theappendixc}{\Alph{appendixc}}
        \renewcommand{\theequation}{\Alph{appendixc}.\arabic{equation}}
        \noindent{\bf Appendix \theappendixc #1}\par\vspace*{0.4cm}}

\renewenvironment{thebibliography}[1]
    {\begin{list}{\arabic{enumi}.}
    {\usecounter{enumi}\setlength{\parsep}{0pt}
\setlength{\leftmargin 1.25cm}{\rightmargin 0pt}
     \setlength{\itemsep}{0pt} \settowidth
    {\labelwidth}{#1.}\sloppy}}{\end{list}}
\newcounter{itemlistc}
\newcounter{romanlistc}
\newcounter{alphlistc}
\newcounter{arabiclistc}

\newcommand{\fcaption}[1]{
        \refstepcounter{figure}
        \setbox\@tempboxa = \hbox{\footnotesize Fig.~\thefigure. #1}
        \ifdim \wd\@tempboxa > 6in
           {\begin{center}
        \parbox{6in}{\footnotesize\baselineskip=15pt Fig.~\thefigure. #1}
            \end{center}}
        \else
             {\begin{center}
             {\footnotesize Fig.~\thefigure. #1}
              \end{center}}
        \fi}
\newcommand{\tcaption}[1]{
        \refstepcounter{table}
        \setbox\@tempboxa = \hbox{\footnotesize Table~\thetable. #1}
        \ifdim \wd\@tempboxa > 6in
           {\begin{center}
        \parbox{6in}{\footnotesize\baselineskip=15pt Table~\thetable. #1}
            \end{center}}
        \else
             {\begin{center}
             {\footnotesize Table~\thetable. #1}
              \end{center}}
        \fi}
 1 
\font\twelverm=cmr10  scaled\magstep 1  
 1  
 1  
 1  
 1  
 1  

\font\tenrm=cmr10

\font\ninerm=cmr9

\evensidemargin=\oddsidemargin  
\def\today{\number\day
           \space\ifcase\month\or
             January\or February\or March\or April\or May\or June\or
             July\or August\or September\or October\or November\or December\fi
           \space\number\year}
\begin{document}
\vspace{1.7cm}
\thispagestyle{empty}
\begin{flushright}
hep-ph/9609326  \hfill {\small DESY}-96-079 \\[-0.3cm]
September,1996  \hfill {\small FERMILAB-PUB}-96/048-{\small T}
\end{flushright}

\vspace{0.9cm}
\centerline{\large\bf 
The Equivalence Theorem And Its Radiative-Correction-Free}
\baselineskip=17pt
\centerline{\large\bf Formulation For All ${\boldmath R_\xi}$ Gauges}

\vspace*{1.1cm}
\centerline{\normalsize 
          {\bf  Hong-Jian He}~~$^{(a)}$\footnote{
                  Electronic address: HJHE@DESY.DE } 
~~~{\small and}~~~ {\bf  William B. Kilgore}~~$^{(b)}$\footnote{
                Electronic address: kilgore@fnal.gov }  }
\vspace*{0.7cm}
\baselineskip=17pt
\baselineskip=17pt
\centerline{\normalsize\it
$^{(a)}$~Theory Division, Deutsches Elektronen-Synchrotron DESY}
\centerline{\normalsize\it D-22603 Hamburg, Germany}
\centerline{\normalsize \it
$^{(b)}$~Department of Theoretical Physics, 
Fermi National Accelerator Laboratory}
\centerline{\normalsize\it P.O. Box 500, Batavia, IL 60510, USA}

\vspace{-1cm}
\thispagestyle{empty}
\renewcommand{\baselinestretch}{0.9}
\begin{abstract}
\begin{center}
{\bf Abstract}
\end{center}
\noindent
{\small
The electroweak equivalence theorem quantitatively connects 
the physical amplitudes of longitudinal massive gauge bosons
to those of the corresponding {\it unphysical} would-be Goldstone bosons. 
Its precise form depends on both the gauge
fixing condition and the renormalization scheme.  
Our previous modification-free schemes have applied to a broad class of
$R_\xi$-gauges including 't~Hooft-Feynman gauge
but excluding Landau gauge.  In this paper we
construct a new renormalization scheme in which the radiative
modification factor, $C_{\rm mod}^a$~, is equal to unity for all
$R_\xi$-gauges, including both 't~Hooft-Feynman and Landau gauges.
This scheme makes $C_{\rm mod}^a$ equal to unity by specifying
a convenient subtraction condition for the would-be Goldstone boson
wavefunction renormalization constant $Z_{\phi^a}$.  
We build the new scheme for both the standard model and the
effective Lagrangian formulated electroweak theories (with either
linearly or non-linearly realized symmetry breaking sector).
Based upon these, a new prescription, called `` divided equivalence
theorem '', is further proposed for extending the high energy region
applicable to the equivalence theorem.

\noindent
PACS number(s): 12.15.Ji, 11.10.Gh, 11.30.Qc, 11.15.Ex} \hfill
(Phys.Rev.D, in press)

\begin{flushright}
\end{flushright}
\end{abstract}

\def\ie{\hbox{\it i.e.}}        \def\etc{\hbox{\it etc.}}
\def\eg{\hbox{\it e.g.}}        \def\cf{\hbox{\it cf.}}
\def\etal{\hbox{\it et al.}}

\def\dm{\partial^{\vphantom\mu}_{\mu}}
\def\dM{\partial_{\vphantom\mu}^{\mu}}

\def\SULY{$SU(2)_L\otimes U(1)_Y$}
\def\ET{equivalence theorem}
\def\GB{Goldstone boson}

\def\bra#1{\left\langle #1\right|}
\def\ket#1{\left| #1\right\rangle}
\def\bracket#1{\VEV{\left|#1\right|}}
\def\bracketc#1{\VEV{\left|#1\right|}_{c}}
\def\braket#1#2{\VEV{#1 | #2}}
\def\mtrxelm#1#2#3{\left\langle#1\left|#2\right|#3\right\rangle}
\def\VEV#1{\left\langle #1\right\rangle}

\def\TP#1{T\!\left(\! #1 \right)}
\def\ketc#1{\left| #1\right\rangle_{c}}
\def\nket#1{\left| #1\right\rangle}
\def\dbrackl{\left[\mkern-5mu\left[}
\def\dbrackr{\right]\mkern-5mu\right]}
\def\Dbrack#1{\dbrackl#1\dbrackr}

\newcount\figno \global\figno=1
\def\sfig#1{\nfig}
\def\nfig#1#2{\xdef#1{\the\figno}
          {\par\baselineskip=14pt\tenrm
          \centerline{\twelverm Figure\ \the\figno. }
\global\advance\figno by1\begin{quote}#2\end{quote}}}
\global\newcount\secno \global\secno=0
\global\newcount\meqno \global\meqno=1
\def\Ack{\bigbreak\bigskip
\noindent{\bf Acknowledgements}\par\nobreak\medskip\nobreak}

\def\DOE{Fermilab is operated by Universities Research Association,
Inc., under contract DE-AC02-76CH03000 with the United States
Department of Energy.}

\def\newsec#1{\global\advance\secno by1\message{(\the\secno. #1)}
\global\subsecno=0\eqnres@t\noindent{\bf\the\secno. #1}
\par\nobreak\medskip\nobreak}
\def\eqnres@t{\xdef\secsym{\the\secno.}\global\meqno=1\bigbreak\bigskip}
\def\sequentialequations{\def\eqnres@t{\bigbreak}}\xdef\secsym{}
\global\newcount\subsecno \global\subsecno=0
\def\subsec#1{\global\advance\subsecno by1\message{(\secsym\the\subsecno. #1)}
\ifnum\lastpenalty>9000\else\bigbreak\fi
\noindent{\it\secsym\the\subsecno. #1}\par\nobreak\medskip\nobreak}
\def\appendix#1#2{\global\meqno=1\global\subsecno=0\xdef\secsym{\hbox{#1.}}
\bigbreak\bigskip\noindent{\bf Appendix #1. #2}\message{(#1. #2)}
\par\nobreak\medskip\nobreak}
%
%
%
\def\eqnn#1{\xdef #1{(\secsym\the\meqno)}
\global\advance\meqno by1\wrlabeL#1}
\def\eqna#1{\xdef #1##1{\hbox{$(\secsym\the\meqno##1)$}}
\global\advance\meqno by1}
\def\eqn#1#2{\xdef #1{(\secsym\the\meqno)}
\global\advance\meqno by1$$#2\eqno#1$$}

\newskip\kentering \kentering=0pt plus 1000pt minus 1000pt

\newdimen\z@ \z@=0pt 
\newskip\z@skip \z@skip=0pt plus0pt minus0pt
\def\m@th{\mathsurround=\z@}

\newif\ifdt@p
\def\displ@y{\global\dt@ptrue\openup\jot\m@th
  \everycr{\noalign{\ifdt@p \global\dt@pfalse
      \vskip-\lineskiplimit \vskip\normallineskiplimit
      \else \penalty\interdisplaylinepenalty \fi}}}
\def\@lign{\tabskip\z@skip\everycr{}} 

\def\eqalignno#1{\displ@y \tabskip\kentering
  \halign to\displaywidth{\hfil$\@lign\displaystyle{##}$\tabskip\z@skip
    &$\@lign\displaystyle{{}##}$\hfil\tabskip\kentering
    &\llap{$\@lign##$}\tabskip\z@skip\crcr
    #1\crcr}}

\catcode`\@=12

\setcounter{footnote}{00}
\renewcommand{\thefootnote}{\alph{footnote}}
\renewcommand{\baselinestretch}{1.2}
\newpage

\setcounter{page}{1}

\section{Introduction}

The electroweak equivalence theorem (ET)~\cite{et0}-\cite{mike}
quantitatively connects the high energy scattering amplitudes of
longitudinally polarized weak gauge bosons ($V^a_L=W^\pm_L,Z_L$) to
the corresponding amplitudes  of would-be Goldstone bosons 
($\phi^a=\phi^\pm ,\phi^0$).   
The ET has been widely used and has proven to be a powerful tool in
studying the electroweak symmetry breaking (EWSB) mechanism, which
remains a mystery and awaits experimental exploration at
the CERN Large Hadron Collider (LHC) and the future linear colliders.

After some initial proposals~\cite{et0}, Chanowitz and
Gaillard~\cite{mike-mary} gave the first general formulation  of the ET
for an arbitrary number of external longitudinal vector bosons and
pointed out the non-trivial cancellation of terms growing like powers
of the large energy which arise from external longitudinal
polarization vectors. 
The existence of radiative modification factors to the ET 
was revealed by Yao and Yuan and further discussed by 
Bagger and Schmidt~\cite{YY-BS}.
In  recent systematic investigations,
the precise formulation for the ET has been given for both
the standard model (SM)~\cite{et1,et2,bill} 
and chiral Lagrangian formulated electroweak theories (CLEWT)~\cite{et3}, 
in which convenient renormalization schemes for exactly simplifying 
these modification factors have been proposed
for a class of $R_\xi$-gauges. 
A further general study of both multiplicative and additive modification 
factors [cf. eq.~(1.1)] has been performed 
in Ref.~\cite{et4,et5} for both the SM and CLEWT, 
by analyzing the longitudinal-transverse ambiguity 
and the physical content of the ET as a criterion for 
probing the EWSB sector. According to these studies,
the ET can be precisely formulated as~\cite{mike-mary}-\cite{et5}
$$                                                                             
T[V^{a_1}_L,\cdots ,V^{a_n}_L;\Phi_{\alpha}]                                   
= C\cdot T[i\phi^{a_1},\cdots ,i\phi^{a_n};\Phi_{\alpha}]+ B ~~,
\eqno(1.1)                                          
$$                                                                             
$$                                                                             
\begin{array}{ll}   
C & \equiv C^{a_1}_{\rm mod}\cdots C^{a_n}_{\rm mod} 
         = 1 + O({\rm loop}) ~~,\\[0.25cm]
B & \equiv\sum_{l=1}^n (~C^{a_{l+1}}_{\rm mod}\cdots C^{a_n}_{\rm mod}
T[v^{a_1},\cdots ,v^{a_l},i\phi^{a_{l+1}},\cdots ,
i\phi^{a_n};\Phi_{\alpha}]
+ ~{\rm permutations ~of}~v'{\rm s ~and}~\phi '{\rm s}~) \\[0.25cm]
& = O(M_W/E_j){\rm -suppressed}\\[0.25cm]
& v^a\equiv v^{\mu}V^a_{\mu} ~,~~~
v^{\mu}\equiv \epsilon^{\mu}_L-k^\mu /M_V = O(M_V/E)~,~~(M_V=M_W,M_Z)~~,
\end{array}
\eqno(1.1a,b,c)                                         
$$ 
with the conditions
$$
E_j \sim k_j  \gg  M_W ~, ~~~~~(~ j=1,2,\cdots ,n ~)~~,
\eqno(1.2a)
$$
$$
C\cdot T[i\phi^{a_1},\cdots ,i\phi^{a_n};\Phi_{\alpha}] \gg B~~,
\eqno(1.2b)                                         
$$
where $~\phi^a$~'s  are the Goldstone boson fields and 
$\Phi_{\alpha}$ denotes other possible physical in/out states. 
~$C_{\rm mod}^a=1+O({\rm loop})$~ is a 
renormalization-scheme and gauge dependent 
constant called the modification factor, and
$E_j$ is the energy of the $j$-th external line. For $~E_j \gg M_W~$,
the $B$-term is only $~O(M_W/E_j)$-suppressed relative to
the leading term~\cite{et4},
$$
B = O\left(\frac{M_W^2}{E_j^2}\right)
  T[ i\phi^{a_1},\cdots , i\phi^{a_n}; \Phi_{\alpha}] +   
  O\left(\frac{M_W}{E_j}\right)T[ V_{T_j} ^{a_{r_1}}, i\phi^{a_{r_2}},
                      \cdots , i\phi^{a_{r_n}}; \Phi_{\alpha}]~~.
\eqno(1.3)                         
$$ 
Therefore it can be either larger or smaller than 
$~O(M_W/E_j)$, depending on the magnitudes of the $\phi^a$-amplitudes 
on the RHS of (1.3)~\cite{et4,et5}. 
For example, in the CLEWT, it was found that
$~B=O(g^2)~$ \cite{et4,et5,mike}, which is a constant
depending only on the SM gauge coupling constant and
does not vanish with increasing energy.
Thus, the condition (1.2a) is necessary but not sufficient for
ignoring the whole $B$-term. For sufficiency, the condition (1.2b)
must also be imposed~\cite{et4}.
In section~3.3, we shall discuss minimizing the approximation from 
ignoring the $B$-term when going beyond  lowest order calculations.

In the present work, we shall primarily study the simplification of
the radiative modification factors, $C^a_{\rm mod}$~, to unity.
As shown in (1.1), the modification factors differ from unity 
at loop levels for all external would-be
Goldstone bosons, and are not suppressed by the $~M_W/E_j$-factor.  
Furthermore, these modification factors 
may depend on both the gauge and scalar coupling
constants~\cite{et1,et2}.  
Although $~~C^a_{\rm mod}-1 = O({\rm loop})~~$,
this does {\it not} mean that $~C^a_{\rm mod}$-factors cannot appear 
at the leading order of a perturbative expansion. An example
is the $~1/{\cal N}$-expansion~\cite{1/N} 
in which the leading order contributions
include an infinite number of Goldstone boson loops so that 
 the  $~C^a_{\rm mod}$~'s will survive the large-${\cal N}$ limit 
if the renormalization scheme is not properly chosen.  
In general, the appearance of $C^a_{\rm mod}$~'s
at loop levels alters the
high energy equivalence between $V_L$ and Goldstone boson amplitudes 
and potentially invalidates the na\"{\i}ve intuition gained from tree
level calculations.  For practical applications of the ET at loop
levels, the modification factors complicate the calculations and
reduce the utility of the equivalence theorem.  Thus, the
simplification of $C_{\rm mod}^a$ to unity is very useful.

The factor $C_{\rm mod}^a$ has been derived in the general
$R_\xi$-gauges for both the SM~\cite{et1,et2} and CLEWT~\cite{et3},
and been simplified to unity in a renormalization scheme,
called {\it Scheme-II}~ in those references, for a
broad class of $R_\xi$-gauges.  {\it Scheme-II}~ is particularly
convenient for 't~Hooft-Feynman gauge, but cannot be applied to
Landau gauge.  In the present work, we make a natural generalization
of our formalism and construct a new scheme, which we call {\it
Scheme-IV}~, for {\it all} ~$R_\xi$-gauges including both
't~Hooft-Feynman and Landau gauges.  
In the Landau gauge, the exact simplification of $C_{\rm mod}^a$ is
straightforward for the $U(1)$ Higgs theory~\cite{et2,bill}; 
but, for the realistic non-Abelian
theories (such as the SM and CLEWT) the situation is much
more complicated. Earlier Landau gauge formulations of the non-Abelian
case relied on explicit calculation of new loop level quantities,
$~\Delta^a_i$, involving the Faddeev-Popov ghosts~\cite{et2,PHD}.

This new {\it Scheme-IV}~ proves particularly convenient for
Landau gauge. This is very useful since Landau gauge has been widely used
in the literature and proves particularly convenient for studying
dynamical EWSB. For instance, in the CLEWT, the complicated non-linear
Goldstone boson-ghost interaction vertices from the Faddeev-Popov term
(and the corresponding higher dimensional counter terms) vanish in
Landau gauge, while the Goldstone boson and ghost fields remain
exactly massless~\cite{app}.

In the following analysis, we shall adopt the notation of
references~\cite{et1,et2} unless otherwise specified. This paper
is organized as follows:  In section~2 we derive the necessary 
Ward-Takahashi (WT) identities and construct our new renormalization 
scheme.
In section~3, we derive the precise formulation of 
{\it Scheme-IV}~ such that the ET is 
free from radiative modifications (i.e., $C^a_{\rm mod}=1$) in all
$R_\xi$-gauges 
including both Landau and 't~Hooft-Feynman gauges.
This is done for a variety of models including the $SU(2)_L$ Higgs
theory, the full SM, and both the linearly and non-linearly realized 
CLEWT. We further propose a convenient new
prescription, called the `` Divided Equivalence Theorem '' (DET), for
minimizing the error caused by ignoring the $B$-term.
Finally, we discuss the relation of {\it Scheme-IV}~ to our previous
schemes.
In section~4, we perform explicit one-loop calculations to demonstrate
our results.  Conclusions are given in section~5.

\section{The Radiative Modification Factor $C_{\rm mod}^a$
         and Renormalization {\it Scheme-IV}}

In the first part of this section, 
we shall define our model and briefly explain how
the radiative modification factor to the ET ($C_{\rm mod}^a$) originates
from the quantization and renormalization procedures.
Then, we analyze the properties of the $C_{\rm mod}^a$ in different
gauges and at different loop levels. 
This will provide the necessary preliminaries
for our main analyses and make this paper self-contained.
In the second part of this section,  using  WT identities, 
we construct the new renormalization {\it Scheme-IV}~
for the exact simplification of the $C_{\rm mod}^a$-factor
in {\it all} $R_\xi$-gauges including both 
't~Hooft-Feynman and Landau gauges. Our prescription for obtaining
$C_{\rm mod}^a=1$~ {\it does not require any explicit calculations
beyond those needed for the usual on-shell renormalization program.}

\subsection{The Radiative Modification Factor $C_{\rm mod}^a$}

For simplicity, we shall first derive our results in the 
$SU(2)_L$ Higgs theory by taking $~g^{\prime}=0~$ in the electroweak
$SU(2)_L\otimes U(1)_Y$ standard model (SM).
The generalizations to the full SM and to
the effective Lagrangian formulations 
are straightforward (though there are some further complications)
and will be given in later sections. 
The field content for the $SU(2)_L$ Higgs theory
consists of the physical fields, $H$, $W^a_\mu$, and $f$($\bar{f}$)
representing the Higgs, the weak gauge bosons and the fermions, 
respectively, and the unphysical
fields $\phi^a$, $c^a$, and $\bar{c}^a$, representing the would-be
Goldstone bosons, the Faddeev-Popov ghosts, and the anti-ghosts
respectively.  
We quantize the theory using the following general
$R_\xi$-gauge fixing condition
$$
\begin{array}{l}    
\displaystyle
    {\cal L}_{\rm GF} ~=~ -{1\over2}(F^a_0)^2 ~~,\\[0.4cm]
    F^a_0\ ~=~ (\xi_0^a)^{-{1\over2}}\partial_\mu W^{a\mu}_0
	-(\xi_0^a)^{1\over2}\kappa_0^a\phi^a_0 
       ~=~ (\underline{\bf K}_0^a)^T \underline{\bf W}_0^a ~~,\\[0.3cm]
    \underline{\bf K}_0^a \equiv 
    \displaystyle\left( (\xi_0^a)^{-\frac{1}{2}}\partial_{\mu}, 
     -(\xi_0^a)^{\frac{1}{2}}\kappa_0^a \right)^T ~~,~~~
     \underline{\bf W}_0^a \equiv (W_0^{a\mu}, \phi_0^a)^T ~~, 
\end{array}
\eqno(2.1)    
$$
where the subscript ``$_0$'' denotes bare quantities.  
For the case of the $SU(2)_L$ theory, we can
take $~\xi_0^a=\xi_0~$, $~\kappa_0^a =\kappa_0~$, for $a=1,2,3$.
The quantized bare Lagrangian for the $SU(2)_L$ model is
$$
{\cal L}_{SU(2)_L} = -{1\over4}W^{a\mu\nu}_{0}W^{a}_{0\mu\nu}
	+ \left|D_0^\mu\Phi_0\right|^2 - U_0\left(\left|\Phi_0
	\right|^2\right) - {1\over2}(F^a_0)^2 + ({\xi_0^a})^{1\over2}
	\bar{c}^a_0\hat{s}F^a_0 + ~{\cal L}_{\rm fermion}
\eqno(2.2)    
$$
where $~\hat{s}~$ is the Becchi-Rouet-Stora-Tyutin (BRST)~\cite{BRST} 
transformation operator. Since our analysis and formulation
of the ET do not rely on any details of the Higgs potential 
or the fermionic part, we do not list their explicit forms here.

The Ward-Takahashi (WT) and Slavnov-Taylor (ST) identities of 
a non-Abelian gauge theory are most conveniently
derived from the BRST symmetry of the quantized action.
The transformations of the bare fields are
$$
\begin{array}{ll}
  \hat{s} W^{a\mu}_0 = D_0^{a\mu}{c}_0^{a} = \dM{c}_0^{a}
     + g_0\varepsilon^{abc}\!\Dbrack{W^{\mu b}_0{c}_0^{c}}~, &
  \hat{s} H_0 = -\displaystyle {g_0\over2}\!\Dbrack{\phi_0^a{c}_0^{a}
     \vphantom{W^{b}}} ~,\\[0.4cm]
  \hat{s}\phi_0^a = D_0^{\phi}c_0^a 
      = M_{W0}{c}_0^a + \displaystyle{g_0\over2}
     \Dbrack{H^{\phantom b}_0{c}_0^a} + \displaystyle{g_0\over2}
     \varepsilon^{abc}\!\Dbrack{\phi_{0}^{b}{c}_0^c} ~,~~ &
  \hat{s}{c}_0^a = -\displaystyle{g_0\over2}\varepsilon^{abc}\!\Dbrack{
     {c}_{0}^{b}{c}_{0}^{c}} ~, \\[0.4cm]
  \hat{s}F_0^a = \xi_0^{-{1\over2}}\cdot\partial_{\mu}\hat{s}W_0^{a\mu} -
     \xi_0^{1\over2}\kappa_0 \cdot \hat{s}\phi_0^a ~, &
  \hat{s}\bar{c}_0^{a} = -\xi_0^{-{1\over2}}F^{a}_0 ~, 
\end{array}
\eqno(2.3)   
$$
where expressions such as $\Dbrack{W_{0}^{\mu b}c_0^c}\!(x)$ indicate
the local composite operator fields formed from $W_{0}^{\mu b}(x)$ and
$c_0^c(x)$.

The appearance of the modification factor $C_{\rm mod}^a$ to the ET is
 due to the amputation and the renormalization of  external
massive gauge bosons and their corresponding Goldstone boson fields. 
For the amputation, we need a general ST identity for the propagators
of the gauge boson, Goldstone boson 
and their mixing~\cite{YY-BS}-\cite{et2}.
By introducing the external source term
$~~\int dx^4 [J_i\chi_0^i + \bar{I}^ac_0^a +\bar{c}^a_0I^a]~~$
(where $~\chi_0^i~$ denotes any possible fields except the (anti-)ghost
fields) to the generating functional, we get
the following generating equation for connected Green functions:
$$
0=J_i(x)<0|T\hat{s}\chi_i^a(x)|0> -\bar{I}^a(x)<0|T\hat{s}c^a_0(x)|0>
  +<0|T\hat{s}\bar{c}^a_0(x)|0> I^a(x)
\eqno(2.4)      
$$
from which we can derive the ST identity for the matrix propagator of 
$~\underline{\bf W}_0^a~$,
$$
\underline{\bf K}_0^T \underline{\bf D}_0^{ab} (k)
~=~ - \displaystyle
\left[\underline{\bf X}^{ab}\right]^T (k)
\eqno(2.5)     
$$
with 
$$
\underline{\bf D}_0^{ab}(k) 
~=~ <0|T\underline{\bf W}_0^a (\underline{\bf W}_0^b)^T|0>(k) ~~,~~~~
{\cal S}_0(k)\delta^{ab} ~=~ <0|Tc_0^b\bar{c}_0^a|0>(k)~~,
\eqno(2.5a)          
$$
$$
\underline{\bf X}^{ab}(k)
~\equiv ~\hat{\underline{\bf X}}^{ab}(k){\cal S}_0(k)
~\equiv ~\left( 
\begin{array}{l}
\xi_0^{\frac{1}{2}}<0|T\hat{s}W_0^{b\mu}|0> \\
\xi_0^{\frac{1}{2}}<0|T\hat{s}\phi_0^b|0> 
\end{array} \right)_{(k)}\cdot {\cal S}_0(k)~~.
\eqno(2.5b)      
$$
To explain how the modification factor $C_{\rm mod}^a$ to the ET arises, 
we start from the well-known ST identity~\cite{mike-mary}-\cite{et2}
$~~<0|F_0^{a_1}(k_1)\cdots F_0^{a_n}(k_n)\Phi_{\alpha}|0> =0~~$ and
set $~n=1~$, i.e.,
$$
0=G[F_0^a(k);\Phi_{\alpha}]
 =\underline{\bf K}_0^T G[\underline{\bf W}_0^a(k);\Phi_{\alpha}]
 =- [\underline{\bf X}^{ab}]^T T[\underline{\bf W}_0^a(k);\Phi_{\alpha}] ~~.
\eqno(2.6)    
$$
Here $~G[\cdots ]~$ and $~T[\cdots ]~$ denote the Green function and
the $S$-matrix element, respectively. The identity (2.6) leads directly
to
$$
\frac{k_{\mu}}{M_{W0}}T[W_0^{a\mu}(k);\Phi_{\alpha}] 
= \widehat{C}_0^a(k^2)T[i\phi_0^a;\Phi_{\alpha}]
\eqno(2.7)   
$$
with $~\widehat{C}^a_0(k^2)~$ defined as
$$
\widehat{C}^a_0(k^2)\equiv{{1+\Delta^a_1(k^2)+\Delta^a_2(k^2)}\over
	{1+\Delta^a_3(k^2)}} ~~,
\eqno(2.8)   
$$
in which the quantities $\Delta^a_i$ are the proper vertices of the
composite operators
$$
\begin{array}{rcl}
        \Delta^a_1(k^2)\delta^{ab} 
    & = & \displaystyle{g_0\over2M_{W0}}<0|T
	\Dbrack{H^{\phantom b}_0{c}_0^b}|\bar{c}_0^a>(k) ~~,\\[0.4cm]
    \Delta^a_2(k^2)\delta^{ab} 
    & = &  -\displaystyle{g_0\over2M_{W0}}\varepsilon^{bcd}<0|T
	\Dbrack{\phi_{0}^{c}{c}_0^d}|\bar{c}_0^{a}>(k) ~~,\\[0.4cm]
    ik^\mu\Delta^a_3(k^2)\delta^{ab} 
    & = & -\displaystyle{g_0\over2}\varepsilon^{bcd}<0|T
	\Dbrack{W^{\mu b}_0{c}_0^c}|\bar{c}_0^{a}>(k) ~~,
\end{array}
\eqno(2.9)    
$$
which are shown diagrammatically in figure 1.

\vbox{
\par\epsfxsize=\hsize\epsfbox{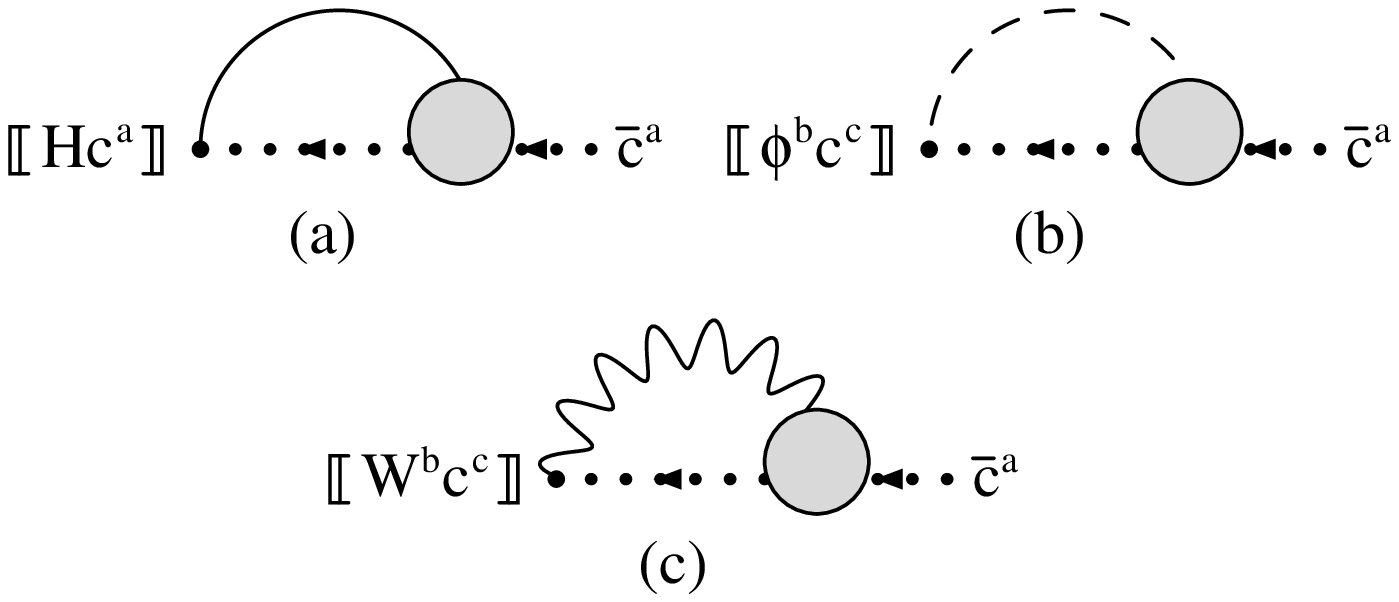}
\sfig{}\figa{Composite operator diagrams contributing to 
radiative modification factor of the equivalence
theorem in non-Abelian Higgs theories. (a). $~\Delta^a_1$~;
(b). $~\Delta^a_2~$; (c). $~\Delta^a_3~$.}}

\noindent
After  renormalization, (2.8) becomes
$$
\frac{k_{\mu}}{M_{W}}T[W^{a\mu}(k);\Phi_{\alpha}] 
= \widehat{C}^a(k^2)T[i\phi^a;\Phi_{\alpha}]
\eqno(2.10)   
$$
with the  finite renormalized coefficient
$$
\widehat{C}^a(k^2) = Z_{M_W}\left(\frac{Z_W}{Z_\phi}\right)^{\frac{1}{2}}
\widehat{C}^a_0(k^2)  ~~.
\eqno(2.11)   
$$
The renormalization constants are defined as
$~W^{a\mu}_0 =Z_W^{1\over 2} W^{a\mu}~$,~
$\phi_0^a = Z_\phi^{1\over 2}\phi^a~$,
and $~M_{W0}=Z_{M_W}M_{W}~$. 
The modification factor to the ET is precisely
the value of this finite renormalized coefficient $~\widehat{C}^a(k^2)~$
on the gauge boson mass-shell:
$$
C^a_{\rm mod} = \left.\widehat{C}^a(k^2)\right|_{k^2=M_W^2}~~,
\eqno(2.12)   
$$
provided that the usual on-shell subtraction for $M_W$ is adopted.
In Sec.~3, we shall transform the identity (2.10) into 
the final form of the ET (which connects the $~W_L^a$-amplitude to
that of the corresponding $\phi^a$-amplitude) 
for an arbitrary number of external longitudinal gauge bosons
and obtain a {\it modification-free} formulation of the ET
with $~C^a_{\rm mod}=1~$ to all loop orders.

As shown above, the appearance of the $C^a_{\rm mod}$ factor to the
ET is due to the amputation and renormalization of external
$W^{a\mu}$ and $\phi^a$ lines by using the ST identity (2.5).
Thus it is  natural that the  $C^a_{\rm mod}$ factor contains
$W^{a\mu}$-ghost, $\phi^a$-ghost and Higgs-ghost interactions expressed
in terms of these $\Delta^a_i$-quantities. Further
simplification can be made by re-expressing $C^a_{\rm mod}$ in terms
of known $W^{a\mu}$ and  $\phi^a$ proper self-energies using
 WT identities as first proposed in Refs.~\cite{et1}-\cite{et3}.
This step is the basis of our simplification of
$~C_{\rm mod}^a =1~$ and will be also adopted 
for constructing our new {\it Scheme-IV}
in  Sec.~2.2. We must emphasize that, {\it our simplification
of $~C_{\rm mod}^a =1~$ does not need any explicit calculation of
the new loop-level $\Delta^a_i$-quantities} which
involve ghost interactions and are quite complicated. This is precisely
why our simplification procedure is useful.

Finally, we analyze the properties of the $~\Delta_i^a$-quantities
in different gauges and at different loop-levels. 
The loop-level $\Delta_i^a$-quantities are 
non-vanishing in general
and make $~\widehat{C}_0(k^2)\neq 1~$ and $~C^a_{\rm mod}\neq 1~$
order by order. In Landau gauge, these $\Delta_i^a$-quantities can be
partially simplified, 
especially at the one-loop order, because the tree-level
Higgs-ghost and $\phi^a$-ghost vertices vanish. This
makes $~\Delta_{1,2}^a=0~$ at one loop.\footnote{ 
We note that, in the non-Abelian case,
the statement that $~\Delta^a_{1,2}=0~$ for Landau gauge
in Refs.~\cite{YY-BS,et2} is only valid at the one-loop order.}~
In general,
$$
\Delta_1^a=\Delta_2^a = 0+O(2~{\rm loop})~,
~~~\Delta_3^a =O(1~{\rm loop})~,~~~~~
(~{\rm in~Landau~gauge}~)~.
\eqno(2.13)        
$$
Beyond the one-loop order, $~\Delta^a_{1,2}\neq 0~$ since the 
Higgs and Goldstone boson fields can still indirectly couple to the ghosts
via loop diagrams containing internal gauge fields, as shown in 
Figure~2.

\vbox{
\par\epsfxsize=\hsize\epsfbox{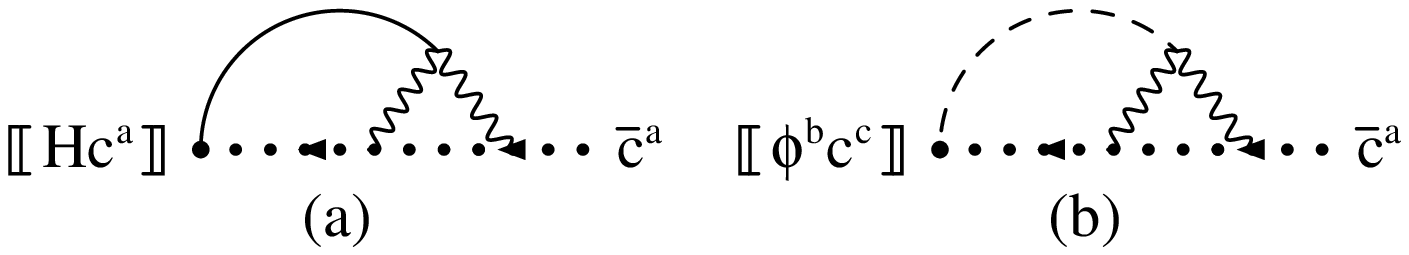}
\sfig{}\figb{The lowest order diagrams contributing to 
$\Delta^a_1(k^2)$ and $\Delta^a_2(k^2)$ in Landau gauge.}}

\noindent
We note that the 2-loop diagram Figure~2b is non-vanishing in the
full SM due to the tri-linear $~A_{\mu}$-$W_{\nu}^{\pm}$-$\phi^{\mp}$ 
and $~Z_{\mu}$-$W_{\nu}^{\pm}$-$\phi^{\mp}$ vertices, while it 
vanishes in the $SU(2)_L$ theory since the couplings of
these tri-linear vertices are proportional to $~\sin^2\theta_W$~.

\subsection{Construction of Renormalization {\it Scheme-IV} }

From the generating equation for WT identities~\cite{et2,BWLee}, 
we obtain a set of identities for bare inverse propagators which
contain the bare modification factor $~\widehat{C}_0^a(k^2)~$ [derived
in (2.8), (2.9) and (2.12)]~\cite{et1,et2}
$$
\begin{array}{rcl}
    ik^\mu[i{\cal D}^{-1}_{0,\mu\nu,ab}(k)+ \xi^{-1}_0 k_\mu k_\nu]\ &=& -
	M_{W0}\widehat{C}^a_0(k^2)[i{\cal D}^{-1}_{0,\phi\nu,ab}(k)- i
	\kappa_0 k_\nu] \\[0.3cm]
    ik^\mu[-i{\cal D}^{-1}_{0,\phi\mu,ab}(k)+i \kappa_0 k_\mu]\ &=& -
	M_{W0}\widehat{C}^a_0(k^2)[i{\cal D}^{-1}_{0,\phi\phi,ab}(k)
	+\xi_0 \kappa^2_0]\\[0.3cm] 
    i{\cal S}^{-1}_{0, ab}(k)\ &=& [1+\Delta_3^a(k^2)][k^2-\xi_0\kappa_0
	M_{W0}\widehat{C}^a_0(k^2)]\delta_{ab}
\end{array}
\eqno(2.14)   
$$
where ${\cal D}_{0,\mu\nu}$, ${\cal D}_{0,\phi\nu}$, ${\cal
D}_{0,\phi\phi}$, ${\cal S}_{0, ab}$ are the unrenormalized full
propagators for gauge boson, gauge-Goldstone-boson mixing and
ghost, respectively.

The renormalization program is chosen to match the on-shell
scheme~\cite{Hollik} for the physical degrees of freedoms, since this
is very convenient and popular for computing the electroweak radiative
corrections (especially for high energy processes).  
Among other things, this choice means that the proper
self energies of physical particles are renormalized so as to vanish
on their mass-shells, and that the vacuum expectation value of the 
Higgs field is renormalized
such that the tadpole graphs are exactly cancelled.  If the vacuum
expectation value were not renormalized in this way, there would be
tadpole contributions to figure~1a.

The renormalization constants of the unphysical degrees of freedoms are
defined as
$$
\phi^a_0 = Z_{\phi}^{1\over2}\phi^a ~,\ \ c^a_0=Z_cc^a ~,
\ \ \bar{c}^a_0=\bar{c}^a ~,\ \ \xi_0^a =Z_\xi \xi^a ~,
\ \ \kappa_0^a =Z_\kappa\kappa^a ~.
\eqno(2.15)   
$$
Some of these  renormalization constants will be chosen such that 
the ET is free from radiative modifications, while the others are left
to be determined as usual~\cite{Hollik} 
so that our scheme is most convenient for the 
practical application of the ET.
Using (2.15) and the relations 
$~{\cal D}_{0,\mu\nu}=Z_W{\cal D}_{\mu\nu},~~
{\cal D}_{0,\phi\nu}=Z^{\frac{1}{2}}_\phi Z^\frac{1}{2}_W
{\cal D}_{\phi \nu}$~, and $~{\cal S}_0 = Z_c{\cal S}~$, we obtain the 
renormalized identities
$$
\begin{array}{l}
    ik^\mu[i{\cal D}^{-1}_{\mu\nu,ab}(k) + 
      {Z_W\over Z_\xi}\xi^{-1}k_\mu
      k_\nu]\ ~=~ - \widehat{C}^a(k^2)M_W[i{\cal D}^{-1}_{\phi\nu,ab}(k)
	-  Z_\kappa Z^{1\over2}_W Z^{1\over2}_\phi ik_\nu\kappa]
     \\[0.35cm]
    ik^\mu[-i{\cal D}^{-1}_{\phi\mu,ab}(k) + Z_\kappa Z^{1\over2}_W
	Z^{1\over2}_\phi ik_\mu\kappa]\ ~=~ -\widehat{C}^a(k^2)M_W[i{\cal
	D}^{-1}_{\phi\phi,ab}(k) + Z^2_\kappa Z_\xi Z_\phi\xi\kappa^2]
     \\[0.35cm]
    i{\cal S}^{-1}_{ab}(k) ~=~ Z_c[1+\Delta_3(k^2)][k^2-\xi\kappa M_W Z_\xi
	Z_\kappa ({Z_\phi\over Z_W})^{1\over2}
	\widehat{C}^a(k^2)]\delta_{ab} 
\end{array}
\eqno(2.16)   
$$
Note that the renormalized coefficient $~\widehat{C}^a(k^2)~$ 
appearing in (2.16) is precisely the same as that in (2.12).

Constraints on $Z_\xi$, $Z_\kappa$, $Z_\phi$ and $Z_c$ 
can be drawn from the fact that the coefficients in the 
renormalized identities of (2.16) are finite. This implies that
$$                                    
\begin{array}{ll}
Z_\xi = \Omega_\xi Z_W~, & Z_\kappa=\Omega_\kappa Z^{\frac{1}{2}}_W
Z^{-\frac{1}{2}}_{\phi}Z_\xi^{-1}~,\\[0.3cm]
Z_\phi =\Omega_\phi Z_W Z^2_{M_W} \hat{C}_0(sub.~ point)~,~~~ &
Z_c=\Omega_c[1+\Delta_3(sub.~ point)]^{-1}~,
\end{array}
\eqno(2.17)             
$$
with
$$
 \Omega_{\xi ,\kappa , \phi , c} ~=~ 1 + O({\rm loop}) ~=~
        {\rm finite}~~,
\eqno(2.17a)            
$$
where $~\Omega_\xi~$, $~\Omega_\kappa~$,
$~\Omega_\phi~$ and $~\Omega_c~$ are unphysical and 
arbitrary finite constants to be determined by the subtraction conditions.

The propagators are expressed in terms of the proper self-energies as

$$
\begin{array}{rcl}
    i{\cal D}^{-1}_{\mu\nu,ab}(k) &=& \left[\left(g_{\mu\nu} -
	{k_\mu k_\nu\over k^2}\right)\left\lgroup -k^2 + M_W^2 -
	\Pi^a_{WW}(k^2)\right\rgroup\right. \\[0.4cm]
	& & \left.\kern30pt + {k_\mu k_\nu\over k^2}\left\lgroup
	-\xi^{-1}k^2 + M_W^2 - {\widetilde\Pi}^a_{WW}(k^2)\right\rgroup
	\right]\delta_{ab} ~~,\\[0.4cm]
    i{\cal D}^{-1}_{\phi\mu,ab}(k) &=& -ik_\mu\left\lgroup M_W -
	\kappa + {\widetilde\Pi}^a_{W\phi}(k^2)\right\rgroup\delta_{ab}~,
       \\[0.3cm]
    i{\cal D}^{-1}_{\phi\phi,ab}(k) &=& \left\lgroup k^2 - \xi\kappa^2
	 - {\widetilde\Pi}^a_{\phi\phi}(k^2)\right\rgroup\delta_{ab}~~,
       \\[0.3cm]
    i{\cal S}^{-1}_{ab}(k) &=& \left\lgroup k^2 - \xi\kappa M_W -
	{\widetilde\Pi}^a_{c\bar{c}}(k^2)\right\rgroup\delta_{ab}~~,
       \\[0.3cm]
\end{array}
\eqno(2.18)   
$$
where $\Pi^a_{WW}$ is the proper self-energy of the physical part of the
gauge boson, and the $\widetilde\Pi^a_{ij}$~'s are the unphysical proper 
self-energies.  

Expanding the propagators in (2.16) in terms of proper self-energies 
yields the following identities containing 
$~\widehat{C}(k^2)~$:
$$
\begin{array}{lll}
    \widehat{C}^a(k^2) & = & 
        \displaystyle{\xi^{-1}k^2\left(\Omega_\xi^{-1} - 1\right) +
	M_W^2 - {\widetilde\Pi}^a_{WW}(k^2)\over
        M_W\kappa\left(\Omega_\kappa\Omega_\xi^{-1}-1\right) +
        M_W^2 + M_W{\widetilde\Pi}^a_{W\phi}(k^2)} ~~,\\[0.5cm]
    \widehat{C}^a(k^2) & = & \displaystyle{k^2\over M_W}{
        \kappa\left(\Omega_\kappa\Omega_\xi^{-1}-1\right)
        + M_W + {\widetilde\Pi}^a_{W\phi}(k^2) 	\over
        \xi\kappa^2\left(\Omega_\kappa^2\Omega_\xi^{-1}-1\right)+	
         k^2 - {\widetilde\Pi}^a_{\phi\phi}(k^2) } ~~,\\[0.5cm]
     {\widetilde\Pi}^a_{c\bar{c}}(k^2) & = &
        k^2 - \xi\kappa M_W - 
	Z_c\left[1+\Delta_3(k^2)\right]\left[M_W^2-\xi\kappa
	\Omega_\kappa M_W \widehat{C}^a(k^2)\right]~~. \\[0.3cm]
\end{array}
\eqno(2.19)       
$$

We are now ready to construct of our new renormalization
scheme, {\it Scheme-IV}~, which will insure 
$~\widehat{C}(M_W^2)=1~$ for all $R_\xi$-gauges, 
including Landau gauge. The $R_\xi$-gauges are a continuous 
one parameter family of gauge-fixing conditions [cf. (2.1)] 
in which the parameter ~$\xi$~ takes values from ~$0$~ to
$~\infty$~.  In practice, however, there are only three important
special cases: the Landau gauge ($\xi=0$), the 't~Hooft-Feynman
gauge ($\xi=1$)  and unitary gauge ($\xi\rightarrow\infty$). In the
unitary gauge, the unphysical degrees of freedom freeze out and
one cannot discuss the amplitude for the would-be
Goldstone bosons. In addition, the loop renormalization
becomes  inconvenient in this gauge 
due to the bad high energy behavior of massive gauge-boson propagators 
and the resulting complication of the divergence structure.
The 't~Hooft-Feynman gauge offers great calculational advantages, 
since the gauge boson propagator
takes a very simple form and the tree-level mass poles of each weak
gauge boson and its corresponding Goldstone boson and ghost are
all the same. The Landau gauge proves very convenient
in the electroweak chiral Lagrangian formalism~\cite{app} 
by fully removing the complicated tree-level
non-linear Goldstone boson-ghost 
interactions [cf. Sec.~3.2] and in this
gauge unphysical would-be Goldstones are exactly massless 
like  true Goldstone bosons.

To construct the new {\it Scheme-IV}~, we note that
{\it a priori}, we have six free parameters to be specified: 
~$\xi$~, ~$\kappa$~, ~$Z_\phi$~, ~$Z_c$~, ~$\Omega_\xi$~, 
and ~$\Omega_\kappa$~ in a general $R_\xi$-gauge.  
For Landau gauge  ($~\xi =0~$), 
the gauge-fixing term $~{\cal L}_{\rm GF}~$ [cf. (2.1)]
gives vanishing Goldstone-boson masses without any $\kappa$-dependence, 
and the bi-linear gauge-boson vertex
$~-\frac{1}{2\xi_0}(\partial_{\mu}W_0^{\mu})^2~$ diverges, implying
that the $W$-propagator is transverse and independent of ~$\Omega_\xi$~.
The only finite term
left in $~{\cal L}_{\rm GF}~$ for Landau gauge is the gauge-Goldstone
mixing vertex
$~~\kappa_0\phi_0\partial_{\mu}W_0^{\mu}
 = ~\Omega_\xi^{-1}\Omega_\kappa\kappa\phi\partial_{\mu}W^{\mu}~~$ 
[cf. (2.17)], 
which will cancel the tree-level $~W$-$\phi$ mixing
from the Higgs kinetic term  $~\left| D_0^{\mu}\Phi_0\right|^2~$ in (2.2)
provided that we choose $~\kappa = M_W~$. Hence, for the purpose
of including Landau gauge into our {\it Scheme-IV}~, we shall not
make use of the degree of freedoms from $\Omega_\xi$ and $\Omega_\kappa$~,
and in order to remove the tree-level $~W$-$\phi~$ mixing, we shall set
$~\kappa = M_W~$.  Thus, we fix the free parameters 
~$\Omega_\xi$~, ~$\Omega_\kappa$ ~and $\kappa$~ as follows
$$
\Omega_\xi ~=~\Omega_\kappa ~=~ 1~~,~~~~   \kappa ~=~ M_W~~,~~~~~
(~{\rm in}~ Scheme-IV~)~~.
\eqno(2.20)      
$$
From (2.17), the choice $~~\Omega_\xi =\Omega_\kappa = 1~~$ implies
$$
F^a_0 = F^a ~~,
\eqno(2.21)        
$$
i.e., the gauge-fixing function $~F^a_0~$ is unchanged after the
renormalization. For the remaining three unphysical parameters 
$~\xi$~, ~$Z_\phi$ ~and ~$Z_c$~, we shall leave ~$\xi$~ free to cover all 
$R_\xi$-gauges and leave ~$Z_c$~ determined by the usual on-shell
normalization condition
$$
{\left. \displaystyle\frac{d}{dk^2}\widetilde{\Pi}^a_{c\bar{c}}
      (k^2) \right| }_{k^2=\xi M_W^2} ~=~0 ~~ .
\eqno(2.22)    
$$
Therefore, in our {\it Scheme-IV}~, {\it the only free parameter,}
which we shall specify for insuring 
$~\widehat{C}(M_W^2)=1~$, is the wavefunction renormalization constant 
$~Z_\phi~$ for the unphysical Goldstone boson.

Under the above choice (2.20),
the first two equations of (2.19) become
$$
\begin{array}{ll}
    \widehat{C}^a(M_W^2) & 
    ~=~ \displaystyle{M_W^2 - {\widetilde\Pi}^a_{WW}(M_W^2)\over M_W^2 +
	M_W{\widetilde\Pi}^a_{W\phi}(M_W^2)}
    ~=~ \displaystyle{M_W^2 + M_W{\widetilde\Pi}^a_{W\phi}(M_W^2)\over
	M_W^2 - {\widetilde\Pi}^a_{\phi\phi}(M_W^2)} \\[0.5cm]
  & ~=~ \displaystyle
        \left[{M_W^2 - {\widetilde\Pi}^a_{WW}(M_W^2)\over M_W^2 -
	{\widetilde\Pi}^a_{\phi\phi}(M_W^2)}\right]^{1\over2} ~~,\\[0.3cm]
\end{array}
\eqno(2.23)     
$$
at $~k^2 = M_W^2~$.
Note that (2.23) re-expresses the factor $~\widehat{C}^a(M_W^2)~$ in terms
of only two renormalized proper self-energies: 
$\tilde{\Pi}_{\phi\phi}$ and  $\tilde{\Pi}_{WW}$ 
(or $\tilde{\Pi}_{W\phi}$ ).  We emphasize that, unlike the most general
relations (2.19) adopted in Refs.~\cite{et1,et2}, 
the identity (2.23) compactly takes the {\it same symbolic form
for any $R_\xi$-gauge including both 't~Hooft-Feynman and Landau gauges}
under the choice (2.20).\footnote{In fact, (2.23) holds for
arbitrary $~\kappa$~.}

From the new identity (2.23), we deduce that the modification factor
$~C^a(M_W^2)~$ can be made equal to unity provided the condition
$$
{\widetilde\Pi}^a_{\phi\phi}(M_W^2)={\widetilde\Pi}^a_{WW}(M_W^2)
\eqno(2.24)     
$$
is imposed.
This is readily done by adjusting $~Z_\phi ~$ in correspondence to the
unphysical arbitrary finite quantity 
$~\Omega_{\phi} = 1+\delta\Omega_{\phi}~$ in (2.17).   
The precise form of the
needed adjustment can be determined by expressing 
the renormalized proper self-energies 
in terms of the bare proper self-energies plus the corresponding counter
terms~\cite{et2}, 
$$
\begin{array}{lll}
    \widetilde{\Pi}^a_{WW}(k^2) & =~ \tilde{\Pi}_{WW ,0}(k^2)
   +\delta\tilde{\Pi}_{WW} & =~ \displaystyle
       Z_W\widetilde{\Pi}^a_{WW,0}(k^2)+
	(1-Z_WZ_{M_W}^2)M_W^2 ~~, \\[0.35cm]
    \widetilde{\Pi}^a_{W\phi}(k^2) & =~ \tilde{\Pi}_{W\phi ,0}(k^2)
   +\delta\tilde{\Pi}_{W\phi} 
    & =~ \displaystyle (Z_WZ_\phi)^{1\over2}
	\widetilde{\Pi}^a_{W\phi,0}(k^2) + [(Z_WZ_\phi
	Z_{M_W}^2)^{1\over2}-1]M_W  ~~, \\[0.35cm]
    \widetilde{\Pi}^a_{\phi\phi}(k^2)  & =~\tilde{\Pi}_{\phi\phi ,0}(k^2)
   +\delta\tilde{\Pi}_{\phi\phi}  & =~ \displaystyle
       Z_\phi\widetilde{\Pi}^a_{
	\phi\phi,0}(k^2) + (1-Z_\phi )k^2 ~~, \\[0.2cm]
\end{array}
\eqno(2.25)    
$$
which, at the one-loop order, reduces to
$$
\begin{array}{ll}
\tilde{\Pi}_{WW}(k^2) & 
   =~\tilde{\Pi}_{WW ,0}(k^2)-[\delta Z_W +2\delta Z_{M_W}]M_W^2~~,\\[0.2cm]
\tilde{\Pi}_{W\phi}(k^2) & 
   =~\tilde{\Pi}_{W\phi ,0}(k^2)
   -[\frac{1}{2}(\delta Z_W +\delta Z_\phi )+ \delta Z_{M_W}]M_W~~,\\[0.2cm]
\tilde{\Pi}_{\phi\phi}(k^2) & 
   =~\tilde{\Pi}_{\phi\phi ,0}(k^2)-\delta Z_\phi k^2 ~~.\\[0.2cm]
\end{array}
\eqno(2.26)     
$$
Note that, in the above expressions for the $R_\xi$-gauge counter terms
under the choice  $~~\Omega_\xi =\Omega_\kappa = 1~~$ [cf. (2.20)], 
there is no explicit dependence on the gauge
parameters $~\xi~$ and $~\kappa~$ so that (2.25) and (2.26) take
the {\it same} forms for all $R_\xi$-gauges.
From either (2.25) or (2.26), we see that
in the counter terms to the self-energies there are three independent  
renormalization constants $~Z_W,~Z_{M_W}$, and ~$Z_\phi~$. Among them,
$~Z_W~$ and $~Z_{M_W}~$ have been determined by the renormalization
of the physical sector, such as in the on-shell scheme (which
we shall adopt in this paper)~\cite{Hollik},
$$
\begin{array}{ll}
\left. \displaystyle\frac{d}{dk^2}\Pi_{WW}^a(k^2)\right|_{k^2=M_W^2}=0~,~~ &
(~{\rm for}~ Z_W~)~;\\[0.58cm]
\Pi_{WW}^a(k^2)|_{k^2=M_W^2}=0~,~~ & (~{\rm for}~ Z_{M_W}~)~.\\[0.3cm]
\end{array}
\eqno(2.27)             
$$
We are just left with $~Z_\phi~$ from the unphysical sector
which can be adjusted, as shown in eq.~(2.17). 
Since the ghost self-energy $~\widetilde{\Pi}^a_{c\bar{c}}~$ is irrelevant
to above identity (2.23), we do not list, in (2.25) and (2.26), 
the corresponding counter term $~\delta\widetilde{\Pi}^a_{c\bar{c}}~$  
which contains one more renormalization constant $~Z_c~$ 
and will be determined as usual [cf. (2.22)].
Finally, note that we have already 
included the Higgs-tadpole counter term $~-i\delta T~$ 
in the bare Goldstone boson and Higgs boson self-energies, through the
well-known ~tadpole~$=0~$ condition~\cite{et2,Hollik,MW}.

Now, equating ${\widetilde\Pi}^a_{\phi\phi}(M_W^2)$ 
and ${\widetilde\Pi}^a_{WW}(M_W^2)$ according to (2.24),
we solve for ~$Z_\phi$:
$$
Z_\phi ~=~ \displaystyle
         Z_W{Z_{M_W}^2M_W^2-\widetilde{\Pi}^a_{WW,0}(M_W^2)\over
	M_W^2-\widetilde{\Pi}^a_{\phi\phi,0}(M_W^2)} ~~,~~~~
        (~{\rm in}~ Scheme-IV~)~.
\eqno(2.28)    
$$
~$Z_\phi~$ is thus expressed in terms of known quantities, that is,
in terms of the renormalization constants of the physical sector and
the bare unphysical proper self-energies of the gauge fields and the
Goldstone boson fields, which must be computed in
any practical renormalization program.  
We thus obtain $~\widehat{C}^a(M_W^2)=1~$ without the extra work of
explicitly evaluating the complicated $\Delta_i^a$~'s. 
At the one-loop level, the solution for $~Z_\phi
= 1 + \delta Z_\phi ~$ in (2.27) reduces to
$$
\delta Z_\phi = 1 + \delta Z_W + 2\delta Z_{M_W}
        + M_W^{-2}\left[\widetilde{\Pi}^a_{\phi\phi,0}(M_W^2)-
	\widetilde{\Pi}^a_{WW,0}(M_W^2)\right] ~~.
\eqno(2.28a)   
$$
If we specialize to Landau gauge, (2.28a) can be alternatively
expressed in terms of the bare ghost self-energy 
$~\widetilde{\Pi}_{c\bar{c},0}~$ plus the gauge boson
renormalization constants:
$$
\delta Z_\phi = 1 + \delta Z_W + 2\delta Z_{M_W}
        + 2M_W^{-2}\widetilde{\Pi}^a_{c\bar{c},0}(M_W^2)~~, 
~~~~~(~\xi =0~)~~,
\eqno(2.28a')   
$$
due to the Landau gauge WT identity (valid up to one loop) 
$$
\widetilde{\Pi}_{\phi\phi,0}(M_W^2)-\widetilde{\Pi}_{WW,0}(M_W^2) 
~=~2\widetilde{\Pi}_{c\bar{c},0}(M_W^2) + ~O(2~{\rm loop}) ~~.
\eqno(2.29)
$$
The validity of (2.29) can be proven directly. From the first two identities 
of our (2.14) we derive
$$
\begin{array}{ll}
\displaystyle
    \widehat{C}_0^a(M_{W0}^2)  & ~=~ \displaystyle
  \left[{M_{W0}^2 - {\widetilde\Pi}^a_{WW,0}(M_{W0}^2)\over M_{W0}^2 -
  {\widetilde\Pi}^a_{\phi\phi ,0}(M_{W0}^2)}\right]^{1\over2} 
 ~~=~\widehat{C}_0^a(M_{W}^2) + ~O(2~{\rm loop})  \\[0.6cm]
 & ~=~ \displaystyle
 1+\frac{1}{2}M_W^{-2}
\left[\widetilde{\Pi}_{\phi\phi,0}(M_W^2)-\widetilde{\Pi}_{WW,0}(M_W^2)\right]
+~O(2~{\rm loop})~~,
\end{array}
\eqno(2.30)     
$$
and from the third identity of (2.14) plus (2.8) and (2.13) we have
$$
\begin{array}{ll}
 \widehat{C}_0^a(M_W^2) \displaystyle & =~
1 - \Delta^a_3(M_W^2) +~O(2~{\rm loop})~~,~~~~~~(~\xi = 0~)~~\\[0.25cm]
& =~1+ M_W^{-2}\widetilde{\Pi}_{c\bar{c},0}^a(M_W^2)
+~O(2~{\rm loop})~~,~~~~~~(~\xi = 0~)~~.
\end{array}
\eqno(3.31)
$$ 
Thus, comparison of (2.30) with (2.31) gives
our one-loop order Landau-gauge WT identity (2.29) 
so that (2.28a$'$) can be simply deduced from (2.28a).
As a consistency check, we note that the same one-loop result
(2.28a$'$) can also be directly derived from (2.8), (2.11) and (2.13)
for Landau gauge by using (3.31) and requiring $~C^a_{\rm mod}=1~$,

In summary, the complete definition of the {\it Scheme-IV}~ for the
$SU(2)_L$ Higgs theory is as follows: The physical sector is
renormalized in 
the conventional on-shell scheme~\cite{et2,Hollik}.  
This means that the vacuum expectation value is renormalized 
so that tadpoles are exactly cancelled, 
the proper self-energies of physical states vanish on their
mass-shells, and the residues of the propagator poles are normalized
to unity.  For the gauge sector, this means that $~Z_W$ ~and ~$Z_{M_W}$~ 
are determined by (2.27).

In the unphysical sector, the parameters ~$\kappa$~, ~$\Omega_\kappa$ 
~and ~$\Omega_\xi$~ are chosen as in (2.20). The ghost wavefunction
renormalization constant $~Z_c~$ is determined as usual [cf. (2.22)].
The Goldstone wavefunction renormalization constant
~$Z_\phi$~ is chosen as in (2.28) [or (2.28a)] 
so that $~~\widehat{C}(M_W^2)=1~~$ is ensured.
From (2.12), we see that this will automatically
render the ET {\it modification-free}.

\subsection{{\it Scheme-IV\/} in the Standard Model}

For the full SM, the renormalization is
greatly complicated due to the various mixings in the neutral 
sector~\cite{et2,Hollik}. However, the first two WT identities in
(2.19) take the {\it same} symbolic 
forms for both the charged  and neutral sectors
as shown in Ref.~\cite{et2}. 
This makes the generalization of our {\it Scheme-IV}~
to the SM  straightforward. Even so, we still
have a further complication in our final result 
for determining the wavefunction renormalization constant 
$~Z_{\phi^Z}~$ of the neutral Goldstone field $~\phi^Z~$, 
due to the mixings in the counter term
to the bare $Z$ boson self-energy. 

The SM gauge-fixing term can be compactly written as follows~\cite{et2}
$$
{\cal L}_{\rm GF} = -\displaystyle{1\over 2}(F_0^+F_0^- + F_0^-F_0^+)
  -{1\over 2}({\bf F}_0^N)^T{\bf F}_0^N ~~,
\eqno(2.32)   
$$
where
$$
\begin{array}{ll}
F^\pm_0 & =~ (\xi_0^\pm)^{-{1\over 2}}\partial_\mu W^{\pm\mu}_0
        -(\xi_0^\pm)^{1\over 2}\kappa_0^\pm\phi^\pm_0 ~~,\\[0.25cm]
{\bf F}_0^N & =~ (F_0^Z, F_0^A)^T 
      = (\xi^N_0)^{-{1\over2}}\partial_\mu {\bf N}_0^\mu 
        -\bar{\kappa}_0\phi_0^Z~~,
\end{array}
\eqno(2.33)  
$$
and
$$
{\bf N}_0^\mu = (Z_0^\mu ,A_0^\mu )^T ~~,~~~~
{\bf N}_0^\mu = {\bf Z}_N^{1\over2}{\bf N}~~;~~~~
 (\xi^N_0)^{-{1\over2}} = (\xi^N)^{-{1\over2}}{\bf Z}_{\xi_N}^{-{1\over2}}~~,
\\[0.5cm]
\eqno(2.34a)
$$
$$
\displaystyle(\xi^N_0)^{-{1\over2}} =\left[
 \begin{array}{ll}  
   (\xi^Z_0)^{-{1\over2}}~ & ~ (\xi^{ZA}_0)^{-{1\over2}}\\[0.2cm]
   (\xi^{AZ}_0)^{-{1\over2}}~ & ~ (\xi^A_0)^{-{1\over2}}\end{array}\right]~~,
~~~~~~
(\xi^N)^{-{1\over2}} =\left[
 \begin{array}{ll}  
   (\xi^Z)^{-{1\over2}}~ & ~ 0 \\[0.2cm]
   0 ~ & ~ (\xi^A)^{-{1\over2}}\end{array}\right]~~;\\[0.5cm]
\eqno(2.34b)
$$
$$
\bar{\kappa}_0 = \left( (\xi^Z_0)^{-{1\over2}}\kappa_0^Z,~
                   (\xi^A_0)^{-{1\over2}}\kappa_0^A\right)^T~,~~~
\bar{\kappa} = 
\displaystyle\left( (\xi^Z)^{-{1\over2}}\kappa^Z,~0 \right)^T~~,~~~
\bar{\kappa}_0 = {\bf Z}_{\bar\kappa} \bar{\kappa} ~~.\\[0.5cm]
\eqno(2.34c)     
$$
The construction of {\it Scheme-IV}~ for
the charged sector is essentially the same as the $SU(2)_L$ theory and will
be summarized below in (2.41). So, we only need to take
care of the neutral sector. We can derive a set of WT identities
parallel to (2.14) and (2.16) as in Ref.~\cite{et2} and obtain the
following constraints on the renormalization constants for
$~\xi_N~$ and $~\bar{\kappa}~$
$$ 
\displaystyle
{\bf Z}_{\xi_N}^{-{1\over2}}
~=~ {\bf \Omega}_{\xi_N}^{-{1\over2}}{\bf Z}_N^{-{1\over2}}~~,~~~~
{\bf Z}_{\bar{\kappa}}= \left(\xi_N^{1\over2}\right)^T
\left[ {\bf \Omega}_{\xi_N}^{-{1\over2}}\right]^T
\left(\xi_N^{-{1\over2}}\right)^T
 {\bf \Omega}_{\bar{\kappa}}Z_{\phi^Z}^{-{1\over2}} ~~,
\eqno(2.35)       
$$
with
$$
\begin{array}{ll}
{\bf \Omega}_{\xi_N}^{-{1\over2}} & ~\equiv~\displaystyle\left[
\begin{array}{ll}
(\Omega_\xi^{ZZ})^{-{1\over2}}~ & (\Omega_\xi^{ZA})^{-{1\over2}}\\[0.3cm]
(\Omega_\xi^{AZ})^{-{1\over2}}~ & (\Omega_\xi^{AA})^{-{1\over2}}
\end{array}\right]~\equiv~\left[
\begin{array}{ll}
(1+\delta\Omega_\xi^{ZZ})^{-{1\over2}}~ & 
-{1\over2}\delta\Omega_\xi^{ZA}\\[0.3cm]
-{1\over2}\delta\Omega_\xi^{AZ}~ & 
(1+\delta\Omega_\xi^{AA})^{-{1\over2}} \end{array}\right] ~~,\\[0.9cm]
{\bf \Omega}_{\bar\kappa} & ~\equiv~ \displaystyle \left[
\begin{array}{ll}
\Omega^{ZZ}_\kappa~ & 0\\[0.3cm]
\Omega^{AZ}_\kappa~ & 0\end{array}\right]~\equiv~\left[
\begin{array}{ll}
1+\delta\Omega_\kappa^{ZZ}~ & 0\\[0.3cm]
\delta\Omega^{AZ}_{\kappa}~ & 0 \end{array} \right] ~~.\\[0.4cm]
\end{array}
\eqno(2.35a,b)   
$$
As in (2.20), we choose
$$
 {\bf \Omega}_{\xi_N} ~=~  \displaystyle\left[
\begin{array}{ll}
1~ & 0 \\[0.2cm]
0~ & 1
\end{array} \right] ~~,~~~~
{\bf \Omega}_{\bar{\kappa}} ~=~ \displaystyle\left[
\begin{array}{ll}
1~ & 0 \\[0.2cm]
0~ & 0  \end{array} \right] ~~,~~~~
 \kappa_Z = M_Z ~~,~~~~(~{\rm in}~ Scheme-IV~)~~.
\eqno(2.36)   
$$

As mentioned above, in the full SM, the corresponding identities for
$~\widehat{C}^W(M_W^2)~$ and $~\widehat{C}^Z(M_Z^2)~$
take the same symbolic forms as (2.23)
$$
\displaystyle
\widehat{C}^W(M_W^2)
~=~ \displaystyle
        \left[{M_W^2 - {\widetilde\Pi}_{W^+W^-}(M_W^2)\over M_W^2 -
	{\widetilde\Pi}_{\phi^+\phi^-}(M_W^2)}\right]^{1\over2} ~~,~~~~
~\widehat{C}^Z(M_Z^2)
~=~ \displaystyle 
        \left[{M_Z^2 - {\widetilde\Pi}_{ZZ}(M_Z^2)\over M_Z^2 -
	{\widetilde\Pi}_{\phi^Z\phi^Z}(M_Z^2)}\right]^{1\over2} ~~,
\eqno(2.37)   
$$
which can be simplified to unity provided that
$$
{\widetilde\Pi}_{\phi^+\phi^-}(M_W^2)~=~{\widetilde\Pi}_{W^+W^-}(M_W^2) 
~~,~~~~ 
{\widetilde\Pi}_{\phi^Z\phi^Z}(M_Z^2)~=~{\widetilde\Pi}_{ZZ}(M_Z^2)~~.
\eqno(2.38)  
$$
The solution for $~Z_{\phi^\pm}~$ from the first condition of (2.38)
is  the same as in (2.28) or (2.28a), but the solution
for $~Z_{\phi^Z}~$ from the second condition of (2.38) is complicated
due to the mixings in the $~\widetilde{\Pi}_{ZZ,0}~$ counter term:
$$
\begin{array}{l}
\widetilde{\Pi}_{ZZ}(k^2) ~ 
=~\widetilde{\Pi}_{ZZ,0}(k^2)+\delta\widetilde{\Pi}_{ZZ}~=~
\displaystyle
       Z_{ZZ}\widehat{\widetilde{\Pi}}_{ZZ,0}(k^2)+
	(1-Z_{ZZ}Z_{M_Z}^2)M_Z^2 ~~, \\[0.2cm]
\displaystyle
~~~~~~\widehat{\widetilde{\Pi}}_{ZZ,0}(k^2)\equiv 
\widetilde{\Pi}_{ZZ,0}(k^2) + 
Z_{ZZ}^{-{1\over2}}Z_{AZ}^{1\over2}
[\widetilde{\Pi}_{ZA,0}(k^2)+\widetilde{\Pi}_{AZ,0}(k^2)]
  + Z_{ZZ}^{-1}Z_{AZ}\widetilde{\Pi}_{AA,0}(k^2) ~~;\\[0.35cm]
   \widetilde{\Pi}_{\phi^Z\phi^Z}(k^2)  =~ \displaystyle
       Z_{\phi^Z}\widetilde{\Pi}_{
	\phi^Z\phi^Z,0}(k^2) + (1-Z_{\phi^Z} )k^2 ~~. \\[0.2cm]
\end{array}
\eqno(2.39)     
$$
Substituting (2.39) into the second condition of (2.38), we find
$$
\begin{array}{ll}
Z_{\phi^Z} & =~ \displaystyle
         Z_{ZZ}{Z_{M_Z}^2M_Z^2-
  \widehat{\widetilde{\Pi}}_{ZZ,0}(M_Z^2)\over
	M_Z^2-\widetilde{\Pi}_{\phi^Z\phi^Z,0}(M_Z^2)} ~~,~~~~
        (~{\rm in}~ Scheme-IV~)~\\[0.5cm]
 & =~1 + \delta Z_{ZZ} + 2\delta Z_{M_Z}
        + M_Z^{-2}\left[\widetilde{\Pi}_{\phi^Z\phi^Z,0}(M_Z^2)-
	\widetilde{\Pi}_{ZZ,0}(M_Z^2)\right] ~~,~~~~
(~{\rm at~}1~{\rm loop}~)~,
\end{array}
\eqno(2.40)    
$$
where the quantity $~\widehat{\widetilde{\Pi}}_{ZZ,0}~$ is defined in
the second equation of (2.39).
The  added complication to the solution
of $~Z_{\phi^Z}~$ due to the mixing effects in the
neutral sector  vanishes at one loop.

Finally, we summarize {\it Scheme-IV}~ for the full SM.
For both the physical and unphysical parts, the
renormalization conditions will be imposed separately for the charged
and neutral sectors.
The conditions for the charged sector are identical to
those for the $SU(2)_L$ theory. In the neutral sector,
for the physical part, the photon and electric charge are renormalized
as in {\it QED}~\cite{Hollik}, while for the unphysical part,
we choose (2.36) and (2.40).
The constraints on the whole unphysical sector in the {\it Scheme-IV}~
are as follows:

$$
\begin{array}{l}
    \displaystyle\hfil\kappa^\pm = M_W~, \hfil
	\Omega_{\xi^\pm} = 1~, \hfil \Omega_{\kappa^\pm} = 1~,\hfil\\[0.5cm]
  \displaystyle  
  \widetilde\Pi_{\phi^+\phi^-}(M_W^2) ~=~ \widetilde\Pi_{W^+W^-}
	(M_W^2) ~~\Longrightarrow ~~ Z_{\phi^\pm} = Z_{W^\pm}{Z_{M_W}^2
	M_W^2-\widetilde{\Pi}_{W^+W^-,0}(M_W^2)\over
	M_W^2-\widetilde{\Pi}_{\phi^+\phi^-,0}(M_W^2)} ~~,\\[0.5cm]
    \displaystyle \delta Z_{\phi^\pm} = \delta
	Z_{W^\pm} + 2\delta Z_{M_W} + M_W^{-2}\left[\widetilde{\Pi}_{
	\phi^+\phi^-,0}(M_W^2) - \widetilde{\Pi}_{W^+W^-,0}(M_W^2)
	\right]~,~~~~(~{\rm at}~1~{\rm loop}~)~;\\[0.3cm]
\end{array}
\eqno(2.41)     
$$
and
$$
\begin{array}{l}
    \displaystyle\hfil\kappa_Z = M_Z~~,~~~~
 {\bf \Omega}_{\xi_N} ~=~  \displaystyle\left[
\begin{array}{ll}
1~ & 0 \\[0.2cm]
0~ & 1
\end{array} \right] ~~,~~~~
{\bf \Omega}_{\bar{\kappa}} ~=~ \displaystyle\left[
\begin{array}{ll}
1~ & 0 \\[0.2cm]
0~ & 0  \end{array} \right] ~~,\\[0.98cm]
  \displaystyle\widetilde\Pi_{\phi^Z\phi^Z}(M_Z^2)~=~ \widetilde\Pi_{ZZ}
	(M_Z^2)~~\Longrightarrow~~ Z_{\phi^Z} = Z_{ZZ}{Z_{M_Z}^2
	M_Z^2-\widehat{\widetilde{\Pi}}_{ZZ,0}(M_Z^2)\over
	M_Z^2-\widetilde{\Pi}_{\phi^Z\phi^Z,0}(M_Z^2)} ~~,\\[0.5cm]
    \displaystyle \delta Z_{\phi^Z} = \delta
	Z_{ZZ} + 2\delta Z_{M_Z} + M_Z^{-2}\left[\widetilde{\Pi}_{
	\phi^Z\phi^Z,0}(M_Z^2) - \widetilde{\Pi}_{ZZ,0}(M_Z^2)
	\right]~,~~~~(~{\rm at}~1~{\rm loop}~)~;
\end{array}
\eqno(2.42)    
$$

\noindent
which insure
$$
    \widehat{C}^W (M_W^2) ~=~ 1~~,
    \hskip 1.5cm
    \widehat{C}^Z(M_Z^2) ~=~ 1~~,~~~~
 (~{\rm in}~ Scheme-IV~)~~.
\eqno(2.43)   
$$
Note that in (2.42) the quantity $~\widehat{\widetilde{\Pi}}_{ZZ,0}~$
is defined in terms of the bare self-energies of the neutral gauge bosons
by the second equation of (2.39) and reduces to 
$~{\widetilde{\Pi}}_{ZZ,0}~$ at one loop.

\section{
Precise Modification-Free Formulation of the ET for All $R_\xi$-Gauges}

In this section, we first give the modification-free formulation 
of the ET within our new {\it Scheme-IV}~ 
for both $~SU(2)_L~$ Higgs theory and the full SM. In Sec.~3.2,
we further generalize our result to the
electroweak chiral Lagrangian (EWCL) formalism~\cite{app,review} 
which provides the most economical description of 
the strongly coupled EWSB sector below the scale of new physics
denoted by the effective cut-off $\Lambda$($\leq 4\pi v\approx 
3.1$~TeV). 
Numerous applications of the ET in this formalism have appeared
in recent years~\cite{LHC}. The generalization to  linearly
realized effective Lagrangians~\cite{linear} 
is much simpler and will be briefly discussed
at the end of Sec.~3.2.  Also, based upon our modification-free
formulation of the ET, we propose
a new prescription, called `` Divided Equivalence Theorem '' (DET),
for minimizing the approximation due to ignoring the additive $B$-term
in the ET.   Finally, in Sec.~3.3, we analyze 
the relation of {\it Scheme-IV}~ to
our previous schemes for the precise formulation of the ET.

\subsection{The Precise Formulation in the $SU(2)_L$ theory and the SM}

From our general formulation in Sec.~2.1, we see that the radiative
modification factor $~C^a_{\rm mod}~$ to the ET is precisely
equal to the factor $~\widehat{C}^a(k^2)~$ evaluated at the physical
mass pole of the longitudinal gauge boson in the usual on-shell scheme.
This is explicitly shown in (2.12) for $SU(2)_L$ theory and
the generalization to the full SM is
straightforward~\cite{et1,et2}
$$
\begin{array}{ll}
C^W_{\rm mod}~=~\widehat{C}^W (M_W^2)~~,~~~~ &
C^Z_{\rm mod}~=~\widehat{C}^Z(M_Z^2)~~,
\end{array}
\eqno(3.1)
$$
for the on-shell subtraction of the gauge boson masses $~M_W~$ and $~M_Z~$.

We then directly apply our renormalization
{\it Scheme-IV} to give a new modification-free formulation of the ET
{\it for all $~R_\xi$-gauges.}
For $SU(2)_L$ Higgs theory, we have
$$
\begin{array}{l}
C^a_{\rm mod}~=~1~~,~~~~~
(~Scheme-IV~{\rm for}~SU(2)_L~{\rm Higgs~theory}~)
\end{array}
\eqno(3.2)
$$
where the {\it Scheme-IV}~ is defined in (2.20) and (2.28,28a).
For the SM, we have
$$
C^W_{\rm mod}~=~1~~,~~~~
C^Z_{\rm mod}~=~1~~,~~~~~(~Scheme-IV~{\rm for}~{\rm SM}~)
\eqno(3.3)
$$
where the {\it Scheme-IV}~ is summarized in (2.41) and (2.42).
We emphasize that {\it the only special step to exactly ensure
$~C^a_{\rm mod}=1~$ and $~C^{W,Z}_{\rm mod}=1~$ is to choose the unphysical
Goldstone boson wavefunction renormalization constants $~Z_\phi~$\/}
as in (2.28) for the $SU(2)_L$ theory and 
$~Z_{\phi^\pm}$ and $~Z_{\phi^Z}~$ as in (2.41)-(2.42) for the SM.

Therefore, we can re-formulate the ET (1.1)-(1.2) in {\it Scheme-IV}~
with the radiative modifications removed:
$$                                                                             
T[V^{a_1}_L,\cdots ,V^{a_n}_L;\Phi_{\alpha}]                                   
= T[i\phi^{a_1},\cdots ,i\phi^{a_n};\Phi_{\alpha}]+ B ~~,
\eqno(3.4)                                          
$$                                                                             
$$                                                                             
\begin{array}{ll}   
B & \equiv\sum_{l=1}^n 
(~T[v^{a_1},\cdots ,v^{a_l},i\phi^{a_{l+1}},\cdots ,
i\phi^{a_n};\Phi_{\alpha}]
+ ~{\rm permutations ~of}~v'{\rm s ~and}~\phi '{\rm s}~) \\[0.25cm]
& = O(M_W/E_j){\rm -suppressed}\\[0.25cm]
& v^a\equiv v^{\mu}V^a_{\mu} ~,~~~
v^{\mu}\equiv \epsilon^{\mu}_L-k^\mu /M_V = O(M_V/E)~,~~(M_V=M_W,M_Z)~,
\end{array}
\eqno(3.4a,b)                                         
$$ 
with the conditions
$$
E_j \sim k_j  \gg  M_W ~, ~~~~~(~ j=1,2,\cdots ,n ~)~,
\eqno(3.5a)
$$
$$
T[i\phi^{a_1},\cdots ,i\phi^{a_n};\Phi_{\alpha}] \gg B ~.
\eqno(3.5b)                                         
$$
Once {\it Scheme-IV}~ is chosen, we need not worry about
 the $C^a_{\rm mod}$-factors
in (1.1)-(1.2) in  {\it any} $~R_\xi$-gauges and to any loop order.
We remark that {\it Scheme-IV}~ is also valid for the 
$1/{\cal N}$-expansion~\cite{1/N} 
since the above formulation is based upon the
 WT identities (for two-point self-energies) 
which take the {\it same} form in any perturbative expansion.
For the sake of many phenomenological applications,
the explicit generalization to the important
effective Lagrangian formalisms will be summarized in the following section.

\subsection{Generalization to the Electroweak Chiral Lagrangian Formalism}

The radiative modification-free formulation of the ET for the
electroweak chiral Lagrangian (EWCL) formalism was given
in Ref.~\cite{et3} for {\it Scheme-II}~ which cannot be used in
 Landau gauge. However, since 
Landau gauge is widely used for the EWCL in the literature
due to its special convenience for this non-linear 
formalism~\cite{app}, 
it is important and useful to generalize our {\it Scheme-IV}
to the EWCL.
As to be shown below, this generalization is straightforward.
We shall summarize our main results for the full
$~SU(2)\otimes U(1)_Y~$ EWCL.
For the convenience of practical applications of the ET
within this formalism, some useful technical details will be provided
in Appendices-A and -B.
In the following analyses, 
we shall not distinguish the notations between 
bare and renormalized quantities unless it is necessary.

We start from the quantized $~SU(2)_L\otimes U(1)_Y~$ bare EWCL
$$
\begin{array}{ll}
{\cal L}^{\rm [q]} & 
=~ {\cal L}_{\rm eff} + {\cal L}_{\rm GF} +{\cal L}_{\rm FP}
\\[0.5cm]
{\cal L}_{\rm eff} & =~\displaystyle
\left[{\cal L}_{\rm G}+{\cal L}^{(2)}+{\cal L}_{\rm F}\right] + 
{\cal L}_{\rm eff}^{\prime}  \\[0.4cm]
{\cal L}_{\rm G} & =~-\displaystyle\frac{1}{2}{\rm Tr}
({\bf W}_{\mu\nu}{\bf W}^{\mu\nu})-\frac{1}{4}B_{\mu\nu}B^{\mu\nu}~~,\\[0.3cm]
{\cal L}^{(2)} & =~\displaystyle\frac{v^2}{4}{\rm Tr}[(D_\mu U)^\dagger
(D^\mu U)] ~~, \\[0.3cm]
{\cal L}_{\rm F} & =~
       \overline{F_{Lj}} i\gamma^\mu D_\mu F_{Lj}
       +\overline{F_{Rj}} i\gamma^\mu D_\mu F_{Rj}
 -(\overline{F_{Lj}}UM_{j} F_{Rj} +\overline{F_{Rj}}M_{j} 
U^{\dagger} F_{Lj} )~~,
\end{array}
\eqno(3.6)    
$$
with
$$
\begin{array}{l}
{\bf W}_{\mu\nu}\equiv\partial_\mu {\bf W}_\nu -\partial_\nu{\bf W}_\mu
                  + ig[{\bf W}_\mu ,{\bf W}_\nu ] ~,~~~~
B_{\mu\nu} \equiv\partial_\mu B_\nu -\partial_\nu B_\mu ~,\\[0.3cm]
U=\exp [i\tau^a\pi^a/v]~~,~~~~
D_\mu U =\partial_\mu U +ig{\bf W}_\mu U -ig^{\prime}U{\bf B}_\mu~~,\\[0.3cm]
\displaystyle
 {\bf W}_\mu \equiv W^a_\mu\frac{\tau^a}{2}~~,~~~
{\bf B}_\mu \equiv B_\mu \frac{\tau^3}{2}~~,~~~\\[0.3cm]
\displaystyle
D_\mu F_{Lj} = \left[\partial_\mu -ig\frac{\tau^a}{2}W^a_\mu-ig^\prime
                  \frac{Y}{2}B_\mu \right] F_{Lj}~~,~~~~
D_\mu F_{Rj} = \left[ \partial_\mu
 -ig^\prime \left(\frac{\tau^3}{2}+\frac{Y}{2}\right)B_\mu
\right] F_{Rj}~~,\\[0.3cm]
F_{Lj}\equiv \left(f_{1j},~f_{2j}\right)_L^T~~,~~~
F_{Rj}\equiv \left(f_{1j},~f_{2j}\right)_R^T~~,
\end{array}
\eqno(3.7)   
$$
where $~\pi^a~$'s are the would-be Goldstone fields in the non-linear
realization; $f_{1j}$ and $f_{2j}$ are the up- and down- type fermions
of the $j$-th family (either quarks or leptons) respectively, 
and all right-handed fermions are $SU(2)_L$ singlet.

In (3.6), the  leading order Lagrangian 
$\left[{\cal L}_{\rm G}+{\cal L}^{(2)}+{\cal L}_{\rm F}\right]$
denotes the model-independent contributions; 
 the model-dependent next-to-leading
order effective Lagrangian $~{\cal L}_{\rm eff}^{\prime}~$ is given
in Appendix-A. 
Many effective operators contained in $~{\cal L}_{\rm eff}^{\prime}~$ 
(cf. Appendix-A), as reflections of the new physics, can be
tested at the LHC and possible future electron (and photon) Linear
Colliders (LC) through longitudinal gauge boson scattering
processes~\cite{et5,LHC,LC}.  Nonetheless, the analysis of the ET and
the modification factors $C^a_{\rm mod}~$ do not depend on the details
of $~{\cal L}_{\rm eff}^{\prime}~$.

The $~SU(2)_L\otimes U(1)_Y~$ 
gauge-fixing term, $~{\cal L}_{\rm GF}$, in (3.6)
is the same as that defined in (2.29) for the SM except that
the linearly realized Goldstone boson fields ($\phi^{\pm ,Z}$) 
are replaced by the non-linearly realized fields ($\pi^{\pm ,Z}$).
The BRST transformations 
for the bare gauge and Goldstone boson fields are
$$
\begin{array}{ll}
\hat{s}W^\pm_\mu & =~-\partial_\mu c^\pm \mp i
   \left[e(A_\mu c^\pm - W^\pm_\mu c^A) 
  + g {\rm c}_{\rm w}(Z_\mu c^\pm - W^\pm c^Z)\right]\\[0.3cm]
\hat{s}Z_\mu & =~ -\partial_\mu c^Z -ig{\rm c}_{\rm w}
  \left[W^+_\mu c^- - W^-_\mu c^+ \right]\\[0.3cm]
\hat{s}A_\mu & =~ -\partial_\mu c^A -ie\left[W^+_\mu c^--W^-_\mu c^+\right]
  \\[0.3cm]
\hat{s}\pi^\pm & =~\displaystyle M_W
 \left[\pm i(\underline{\pi}^Z c^\pm + \underline{\pi}^\pm\tilde{c}^3)
 -\eta\underline{\pi}^\pm 
(\underline{\pi}^+c^-+\underline{\pi}^-c^+)
 -\frac{\eta}{{\rm c}_{\rm w}}\underline{\pi}^\pm 
  \underline{\pi}^Zc^Z
 +\zeta~c^\pm\right] ~~,\\[0.45cm]
\hat{s}\pi^Z & =~ \displaystyle M_Z
\left[i(\underline{\pi}^-c^+-\underline{\pi}^+c^-)-{\rm c}_{\rm w}
\eta\underline{\pi}^Z(\underline{\pi}^+c^-+\underline{\pi}^-c^+)
-\eta\underline{\pi}^Z\underline{\pi}^Zc^Z +  \zeta ~c^Z\right] ~~,
\end{array}
\eqno(3.8)   
$$
where $~{\rm c}_{\rm w} \equiv ~ \cos\theta_{\rm W}~$ and 
$$
\begin{array}{ll}
\tilde{c}^3 & \equiv ~[\cos 2\theta_{\rm W}]c^Z +
                      [\sin 2\theta_{\rm W}]c^A ~~, \\[0.4cm]
\eta & \equiv ~\displaystyle\frac{\underline{\pi}\cot\underline{\pi}-1}
  {\underline{\pi}^2}
    = -\frac{1}{3}+O(\pi^2) ~~,\\[0.4cm]
\zeta & \equiv ~\displaystyle{\underline{\pi}}\cot\underline{\pi}
    = 1 - \frac{1}{3v^2}\vec{\pi}\cdot\vec{\pi}+O(\pi^4)      ~~, \\[0.4cm]
& \underline{\pi}~\equiv~\displaystyle\frac{\pi}{v}~~,~~~~
  \pi ~\equiv~ \left(\vec{\pi}\cdot\vec{\pi}\right)^{\frac{1}{2}} 
     ~=~ \left( 2\pi^+\pi^- +\pi^Z\pi^Z \right)^{\frac{1}{2}} ~~.
\end{array}
\eqno(3.9)   
$$
The derivations for $~\hat{s}\pi^\pm~$ and $~\hat{s}\pi^Z~$ are
given in Appendix-B.
Note that the non-linear realization greatly complicates 
the BRST transformations for the Goldstone boson fields.
This makes the $~\Delta^a_i$-quantities which appear in the
modification factors much more complex.

With the BRST transformations (3.8), we can  write down
the $R_\xi$-gauge Faddeev-Popov ghost Lagrangian in this non-linear 
formalism as:
$$
{\cal L}_{\rm FP} \displaystyle
~=~\xi_W^{1\over2}\left[ \bar{c}^+\hat{s}F^+ + \bar{c}^+\hat{s}F^- \right]
   +\xi_Z^{1\over2} \bar{c}^+\hat{s}F^Z 
   +\xi_A^{1\over2} \bar{c}^+\hat{s}F^A ~~.
\eqno(3.10)   
$$
The full expression is very lengthy due to the complicated
non-linear BRST transformations for $~\pi^a~$'s.
In the Landau gauge, $~{\cal L}_{\rm FP}~$ is greatly simplified
and has the {\it same} form as that in the linearly realized SM,
due to the decoupling of ghosts from the Goldstone bosons at
tree-level. This is clear from (3.10) 
after substituting (2.33) and setting $~~\xi_W=\xi_Z=\xi_A=0~~$
[cf. $(B6)$ in Appendix-B].
This is why the inclusion of Landau gauge into the modification-free 
formulation of the ET is particularly useful.

With these preliminaries, we can now generalize our
precise formulation of the ET to the EWCL formalism. 
In Sec.~2 and 3, our derivation of the factor-$C^a_{\rm mod}~$ and
construction of the renormalization {\it Scheme-IV}~ for simplifying it
to unity are based upon the general ST and WT identities. The validity
of these general identities does {\it not} 
rely on any explicit expression of the $\Delta^a_i$-quantities 
and the proper self-energies, and this makes our generalization 
straightforward. Our results are summarized as follows.

First we consider the derivation of the modification factor-$C^a_{\rm
mod}~$'s from the amputation and renormalization of external $V_L$ and
$\pi$ lines. Symbolically, 
the expressions for $~C^a_{\rm mod}~$'s still have the {\it same}
dependences on the renormalization constants and the $\Delta^a_i$-quantities 
but their explicit expressions are changed in the EWCL
formalism~\cite{et3}.  We consider the charged sector as an example of
the changes.
$$
C^W_{\rm mod}~=~\widehat{C}^W(M_W^2) ~=~\left.
\displaystyle\left(\frac{Z_W}{Z_{\pi^\pm}}\right)^{1\over2}Z_{M_W}
\frac{1+\Delta_1^W (k^2)+\Delta_2^W (k^2)}
     {1+\Delta_3^W (k^2)}\right|_{k^2=M_W^2}
~~,
\eqno(3.11)   
$$
which has the same symbolic form as the linear SM case~\cite{et2} 
[see also (2.18), (2.11) and (2.12) for the $SU(2)_L$ Higgs theory
in the present paper], 
but the expressions for these $~\Delta_i~$'s are given by
$$
\begin{array}{rll}
1+\Delta_1^W (k^2)+\Delta_2^W (k^2) & \equiv & \displaystyle
{1\over M_W}<0|T(\hat{s}\pi^\mp )|\bar{c}^\pm >(k) ~~,\\[0.4cm]
\displaystyle
ik_\mu \left[1+\Delta_3^W (k^2)\right] & \equiv & -<0|T(\hat{s}W^\mp_\mu )
  |\bar{c}^\pm >(k) ~~,\\[0.2cm]
\end{array}
\eqno(3.12)   
$$
where all fields and parameters are bare, and the BRST transformations for 
$~\pi^a~$ and $~W^\pm_\mu~$ are given by (3.8). 
From (3.8)-(3.10) we see that
the expression for $~\Delta_1^W (k^2)+\Delta_2^W (k^2)~$ has been
greatly complicated due to the non-linear transformation of the
Goldstone bosons, while the $~\Delta_3^W~$ takes the same symbolic
form as in the linear SM. For Landau gauge,
these $~\Delta_i~$'s still satisfy the relation (2.13)
and the two-loop graph of the type of Fig.~2b
also appears in the $~\Delta_1^W (k^2)+\Delta_2^W (k^2)~$ of (3.10).
We do not give any further detailed expressions for these
$~\Delta_i~$'s in either charged or neutral sector, because the 
following  formulation of the ET 
within the {\it Scheme-IV}~ does {\it not} rely on any of 
these complicated quantities for $~C^a_{\rm mod}~$.

Second, consider the WT identities derived in Sec.~2, which
enable us to re-express $~C^a_{\rm mod}$ in terms of
the proper self-energies of the gauge and Goldstone fields
and make our construction of the {\it Scheme-IV}~ possible.
For the non-linear EWCL, the main difference is that we now
have higher order effective operators in $~{\cal L}_{\rm eff}^{\prime}~$
(cf. Appendix-A) which
 parameterize the new physics effects below
the effective cutoff scale $~\Lambda~$. 
However, they do not
affect the WT identities for self-energies derived in
Sec.~2 because their contributions, by definition, can always be included
into the bare self-energies, as  was done in Ref.~\cite{app}.
Thus, our the construction for the  {\it Scheme-IV}~ in Sec.~2
holds for the EWCL formalism.  Hence, our final conclusion
on the modification-free formulation of the ET in this formalism
is the same as that given in (3.1)-(3.5) of  Sec.~3.1, after simply
replacing the linearly realized Goldstone boson fields ($\phi^a$~'s)
by the non-linearly realized fields ($\pi^a$~'s).

Before concluding this section, we remark upon another popular
effective Lagrangian formalism~\cite{linear} for the weakly coupled
EWSB sector (also called the decoupling scenario).
In this formalism, the lowest order Lagrangian is just the
linear SM with a relatively light Higgs boson 
and all higher order effective
operators must have dimensionalities larger than four and are
suppressed by the cutoff scale ~$\Lambda$~:
$$
{\cal L}_{\rm eff}^{\rm linear} ~=~\displaystyle 
{\cal L}_{\rm SM}+\sum_{n} \frac{1}{\Lambda^{d_n-4}}{\cal L}_n
\eqno(3.13)   
$$
where $~d_n~(\geq 5)~$ is the dimension of the effective
operator $~{\cal L}_n~$ and the effective cut-off $~\Lambda~$ has, in
principle, no upper bound.
The generalization of our modification-free 
formulation of the ET to this formalism is extremely simple. 
All our discussions in Sec.~2 and 3.1
hold and the only new thing is to put the new physics
contributions to the self-energies into the bare self-energies
 so that the general relations between the bare and
renormalized self-energies [cf. (2.25) and (2.39)] remain the same.
This is similar to the case of the  non-linear EWCL
(the non-decoupling scenario).

\subsection{Divided Equivalence Theorem: a New Improvement}

In this section, for the purpose of minimizing the approximation
from ignoring the additive $B$-term in the ET (3.4) or (1.1),
we propose a convenient new prescription, 
called `` Divided Equivalence Theorem '' (DET), based upon our
modification-free formulation (3.4).

We first note that the rigorous {\it Scheme-IV}~ and the previous 
{\it Scheme-II}~\cite{et1,et2} (cf. Sec.~3.4)
{\it do not rely on the size of the $B$-term}. Furthermore, the result
$~C^a_{\rm mod}=1~$ greatly simplifies the expression for the $B$-term 
[cf. (1.1) and (3.4)]. 
This makes any further exploration and application of either
the physical or technical content of the ET very convenient. 
In the following, we show how the error caused by ignoring $B$-term
in the ET can be minimized through the new prescription DET.

For any given perturbative expansion up to a finite order $N$, 
the $S$-matrix $T$ (involving $V_L^a$
or $\phi^a$) and the $B$-term can be generally written as 
$~~~T=\sum_{\ell =0}^{N}T_{\ell}~~~$ and 
$~~~B=\sum_{\ell =0}^{N}B_{\ell}~~~$. Within our 
modification-free formulation (3.4) of the ET, we have no complication
due to the expansion of each $~C^a_{\rm mod}$-factor on the RHS of (1.1)
[i.e., $~~C^a_{\rm mod}=\sum_{\ell =0}^{N}
\left(C^a_{\rm mod}\right)_{\ell}~~$]. 
Therefore, at $\ell$-th order and with $~C^a_{\rm mod}=1~$ insured,
the exact ET identity in (3.4) can be expanded as
$$
T_{\ell}[V^{a_1}_L,\cdots ,V^{a_n}_L;\Phi_{\alpha}]                  
~~=~~ T_{\ell}[i\phi^{a_1},\cdots ,i\phi^{a_n};\Phi_{\alpha}]
    ~+~ B_{\ell} ~~,
\eqno(3.14)   
$$
and the conditions (3.5ab) become, at the $\ell$-th order,
$$
E_j \sim k_j  \gg  M_W ~, ~~~~~(~ j=1,2,\cdots ,n ~)~,
\eqno(3.15a)
$$
$$
T_{\ell}[i\phi^{a_1},\cdots ,i\phi^{a_n};\Phi_{\alpha}] \gg B_{\ell} ~,
~~~~~(~ \ell =0,1,2,\cdots ~)~.
\eqno(3.15b)   
$$
We can estimate the $\ell$-th order $B$-term as [cf. (1.3)]
$$
B_{\ell} = O\left(\frac{M_W^2}{E_j^2}\right)
  T_{\ell}[ i\phi^{a_1},\cdots , i\phi^{a_n}; \Phi_{\alpha}] +   
  O\left(\frac{M_W}{E_j}\right)
  T_{\ell}[ V_{T_j} ^{a_{r_1}}, i\phi^{a_{r_2}},
                      \cdots , i\phi^{a_{r_n}}; \Phi_{\alpha}]~~,
\eqno(3.16)                         
$$ 
which is $~O(M_W/E_j)$-suppressed for $~~E_j\gg M_W~$.
When the next-to-leading order (NLO: $\ell = 1$~) contributions 
(containing possible new physics effects, cf. Appendix-A for instance) 
are included, the main limitation\footnote{We must clarify that,
for the discussion of the {\it physical content} of the ET as a criterion
for probing the EWSB, as done in Refs.~\cite{et4,et5}, the issue of
including/ignoring the $B$-term is essentially {\it irrelevant} because
both the Goldstone boson amplitude and the $B$-term are explicitly
estimated order by order and are compared to each other~\cite{et4,et5}.}~
on the predication of the ET for the $V_L$-amplitude via computing 
the Goldstone boson amplitude 
is due to ignoring the leading order $B_0$-term. This leading $B_0$-term
is of $O(g^2)$~\cite{et4,et5} in the heavy Higgs SM and the CLEWT and 
cannot always be ignored in comparison with the NLO $\phi^a$-amplitude
$~T_1~$ though we usually have $~T_0\gg B_0~$ and 
$~T_1\gg B_1~$ respectively~\cite{et4,et5} because of
(3.16). It has been shown~\cite{et5} that, 
except some special kinetic regions,
$~~T_0\gg B_0~~$  and $~~T_1\gg B_1~~$ for all effective operators
containing pure Goldstone boson interactions (cf. Appendix-A), 
as long as $~E_j\gg M_W~$~.
Based upon the above new equations (3.14)-(3.16), we can precisely 
formulate the ET at {\it each given order-$\ell$} 
of the perturbative expansion
where only $~B_{\ell}~$, {\it but not the whole $B$-term}, will be
ignored to build the longitudinal/Goldstone boson equivalence.
Hence, {\it the equivalence is divided order by order} in the perturbative
expansion, and the condition for this divided equivalence is
$~~T_{\ell} \gg B_{\ell}~~$ (at the $\ell$-th order)
which is much weaker than $~~T_{\ell}\gg B_0~~$ [deduced from (3.5b)]
for $~\ell\geq 1~$. For convenience, we call this formulation as 
`` Divided Equivalence Theorem '' (DET). 
Therefore, to improve the prediction of $V_L$-amplitude for 
the most interesting NLO contributions (in $~T_1~$) 
by using the ET, we propose the following simple prescription:
\begin{description}
 \item[{\bf (i).}] Perform a direct and precise unitary gauge
calculation for the tree-level $V_L$-amplitude $~T_0[V_L]~$ 
which is quite simple.
\item[{\bf (ii).}]  
Make use of the DET (3.14) and deduce $~T_1[V_L]~$ from the
Goldstone boson amplitude $~T_1[GB]~$, by ignoring $~B_1~$ only. 
\end{description}
To see how simple the direct unitary gauge calculation of the tree-level
$V_L$-amplitude is, we calculate the $~W^+_LW^-_L\rightarrow W^+_LW^-_L~$
scattering amplitude in the EWCL formalism as a typical example. 
The exact tree-level amplitude $~T_0[W_L]~$ only takes three lines:
$$
\begin{array}{l}
T_0[W_L]=
 ig^2\left[ -(1+x)^2 \sin^2\theta 
+ 2x(1+x)(3\cos\theta -1)
-{\rm c}_{\rm w}^2
   \displaystyle\frac{4x(2x+3)^2\cos\theta}{4x+3
  -{\rm s}_{\rm w}^2{\rm c}_{\rm w}^{-2}}\right.\\[0.4cm]
  \left.  +{\rm c}_{\rm w}^2
   \displaystyle\frac{8x(1+x)(1-\cos\theta )(1+3\cos\theta )
   +2[(3+\cos\theta )x+2][(1-\cos\theta )x-\cos\theta ]^2}
   {2x(1-\cos\theta )+{\rm c}_{\rm w}^{-2}} \right]\\[0.4cm]
 +ie^2 \left[ -\displaystyle\frac{x(2x+3)^2\cos\theta}
             {x+1}+4(1+x)(1+3\cos\theta )+
   \displaystyle\frac{[(3+\cos\theta )x+2][(1-\cos\theta )x
              -\cos\theta ]^2}{x(1-\cos\theta )} \right]~,
\end{array}     
\eqno(3.17)     
$$
where $~\theta~$ is the scattering angle, 
$~x\equiv p^2/M_W^2~$ with $~p~$ denoting the C.M. momentum, and
$~{\rm s}_{\rm w}\equiv \sin\theta_W~,
~{\rm c}_{\rm w}\equiv \cos\theta_W~$.
(3.17) contains five diagrams: 
one contact diagram, two $s$-channel diagrams
by $Z$ and photon exchange, and two similar $t$-channel diagrams. 
The corresponding Goldstone boson amplitude also contain five similar
diagrams except all external lines being scalars. However, even for
just including the leading $B_0$-term which contains only one 
external $v^a$-line [cf. (3.4a) or (1.1b)], one has to compute
extra $~5\times 4=20~$ tree-level graphs due to all possible permutations
of the external $v^a$-line. 
It is easy to figure out how the number of these
extra graphs will be greatly increased 
if one explicitly calculates the whole
$B$-term.  Therefore, we point out that, 
as the lowest order tree-level $V_L$-amplitude is 
concerned, it is {\it much simpler} to directly calculate the 
precise tree-level $V_L$-amplitude in the unitary gauge 
than to indirectly calculate 
the $R_\xi$-gauge Goldstone boson amplitude {\it plus} the complicated 
$B_0$ or the whole $B$ term [which contains much more diagrams 
due to the permutations of $~v_\mu$-factors
in (1.1b) or (3.4a)] as proposed in some literature~\cite{other}.
To minimize the numerical error related to the $B$-term, 
our new prescription DET is the best and the most convenient.

Then, let us further exemplify, up to the NLO of the EWCL formalism,
how the precision of the ET is improved by the above new prescription
DET. Consider the lowest order contributions from 
$~{\cal L}_{\rm G}+{\cal L}^{(2)}+{\cal L}_{\rm F}~$ 
[cf. (3.6)] and the NLO contributions
from the important operators $~{\cal L}_{4,5}~$ (cf. Appendix A).
For the typical process $~W_LW_L\rightarrow W_LW_L~$ up to the NLO,
both explicit calculations and the power counting analysis~\cite{et4,et5}
give
$$
\begin{array}{ll}
\displaystyle
T_0 ~=~ O\left(\frac{E^2}{v^2}\right) ~~,~~~~~ &
B_0 ~=~ O(g^2)~~; \\[0.4cm]
\displaystyle 
T_1 ~=~ O\left(\frac{E^2}{v^2}\frac{E^2}{\Lambda^2}\right)~~,~~~~~ &
\displaystyle
B_1 ~=~ O\left(g^2\frac{E^2}{\Lambda^2}\right)~~;
\end{array}
\eqno(3.18)   
$$
where $~v=246$~GeV and $~\Lambda\simeq 4\pi v \simeq 3.1$~TeV.
From the condition (3.5b) and (3.15b) and up to the NLO, we have
$$
\begin{array}{lll}
(3.5b):  & T_1 \gg B_0 ~~\Longrightarrow~~ 1\gg 24.6\%~~, 
& ({\rm for}~ E=1~{\rm TeV})~~;\\[0.3cm]
(3.15b): & T_0 \gg B_0 ~~\Longrightarrow~~ 1\gg 2.56\%~~,
& ({\rm for}~ E=1~{\rm TeV})~~;\\[0.3cm]
         & T_1 \gg B_1 ~~\Longrightarrow~~ 1\gg 2.56\%~~, 
& ({\rm for}~ E=1~{\rm TeV})~~.
\end{array}
\eqno(3.19)  
$$
Here we see that, up to the NLO and for $~E=1$~TeV, 
the precision of the DET (3.14)-(3.16) is increased 
by about a factor of $10$
in comparison with the usual prescription of the ET  [cf. (3.5a,b)]
as ignoring the $B$-term is concerned. 
It is clear that {\it the DET (3.14)-(3.16)
can be applied to a much wider high energy
region than the usual ET due to the much weaker condition (3.15b).}

In general, to do a calculation up to any order $~\ell \geq 1~$,
we can apply the DET to minimize the approximation
due to the $B$-term by following way:
computing the full $V_L$-amplitude up to 
the $(\ell -1)$-th order and applying
the DET (3.14) at $\ell$-th order with $~B_{\ell}~$ ignored.
The practical applications of this DET up to NLO
($\ell =1$) turns out most convenient. 
It is obvious that the above formulation for 
DET generally holds for both the SM and 
the effective Lagrangian formalisms.

\subsection{Comparison of Scheme-IV with Other Schemes}

The fact that we call the new renormalization scheme, {\it Scheme-IV}~,
implies that there are three other previous renormalization
schemes for the ET.  {\it Schemes-I~} and {\it -II}~ 
were defined in references~\cite{et1,et2}, while {\it
Scheme-III}~ was defined in reference~\cite{et4}.

{\it Scheme-I}~\cite{et1,et2} is a generalization of the usual one-loop
level on-shell scheme~\cite{Hollik} to all orders. 
In this scheme, the unphysical sector is renormalized such that, for
example, in the pure $SU(2)_L$ Higgs theory
$$  
\begin{array}{c}
    \widetilde{\Pi}^a_{WW}(\xi\kappa M_W)
        =\widetilde{\Pi}^a_{W\phi}(\xi\kappa M_W)
        =\widetilde{\Pi}^a_{\phi\phi}(\xi\kappa M_W)
        =\widetilde{\Pi}^a_{c\bar{c}}(\xi\kappa M_W)=0~~,
\end{array}
\eqno(3.20a)
$$
$$
\begin{array}{c}
\displaystyle
    \left.{d\over dk^2}\widetilde{\Pi}^a_{\phi\phi}(k^2)\right|_{k^2=
        \xi\kappa M_W} ~=~0~~,~~~~~ 
    \left.{d\over dk^2}\widetilde{\Pi}^a_{c
        \overline{c}}(k^2)\right|_{k^2=\xi\kappa M_W} ~=~ 0~~,
\end{array}
\eqno(3.20b)
$$
where $~k^2=\xi\kappa M_W~$ is the tree level mass pole of the 
unphysical sector.  In this scheme,
the modification factor is not unity, but does take a very simple form
in terms of a single parameter determined by the renormalization 
scheme~\cite{et1,et2},
$$
C^a_{\rm mod} ~=~ \Omega_\kappa^{-1} ~~, ~~~~~ 
(~Scheme-I~{\rm with}~\kappa =M_W~{\rm and}~\xi =1~)~~.
\eqno(3.21)
$$

{\it Scheme-II}~\cite{et1,et2} is a variation of the 
usual on-shell scheme, in which the unphysical sector 
is renormalized such that $C^a_{mod}$ is set equal to unity. 
The choice here is to impose all of
the conditions in (3.20) except that the Goldstone boson wavefunction
renormalization constant $Z_\phi$ is not determined by (3.20b) but
specially chosen. To accomplish this, $\Omega_\xi$ is adjusted so
that ~$\widetilde{\Pi}^a_{WW}(\xi\kappa M_W)=0~$, and $Z_\phi$ is adjusted
so that $\widetilde{\Pi}^a_{\phi\phi}(\xi\kappa M_W)=0$.  $\Omega_\kappa$
is set to unity, which ensures that $~\widetilde{\Pi}^a_{W\phi}(\xi\kappa
M_W)=\widetilde{\Pi}^a_{c\bar{c}}(\xi\kappa M_W)=0~$, and $Z_c$ is
adjusted to ensure that the residue of the ghost propagator is
unity.  The above conditions guarantee that $\widehat{C}^a(\xi\kappa
M_W)=1$.  The final choice is to set $~\kappa = \xi^{-1}M_W~$ so
that $~C^a_{\rm mod}=1~$.  This scheme is particularly convenient for
the  't~Hooft-Feynman gauge, where $\kappa = M_W$.  
For $~\xi\neq 1~$, there is a complication
due to the tree level gauge-Goldstone-boson mixing term
proportional to $~\kappa -M_W = (\xi^{-1}-1)M_W~$. But this is
 not a big problem since the mixing term corresponds to a
tree level gauge-Goldstone-boson propagator similar to that found in
the Lorentz gauge ($~\kappa =0~$)~\cite{lorentz}. The main shortcoming of
 this scheme is that it does not include Landau gauge 
since, for $~\xi =0~$,
the choice $~\kappa = \xi^{-1}M_W~$ is singular 
and the quantities $~\Omega_{\xi ,\kappa}~$ have no meaning.
In contrast to {\it Scheme-II}~, {\it Scheme-IV}~ is
valid for all $R_\xi$-gauges including both Landau and 't~Hooft-Feynman
gauges. The primary inconvenience of {\it Scheme-IV}~ is 
that for non-Landau gauges all unphysical mass poles 
 deviate from their tree-level values~\cite{RT,Hollik,et2},
thereby invalidating condition (3.20a).\footnote{
The violation of (3.20a) in non-Landau gauges
is not special to {\it Scheme-IV}~,
but is a general feature of all schemes~\cite{RT}-\cite{BV} 
which choose the renormalization prescription (2.21) 
for the gauge-fixing condition~\cite{RT,Hollik,et2}.}~ 
This is not really a problem since
these poles have no physical effect.

{\it Scheme-III}~\cite{et4} is specially designed for 
the pure $V_L$-scatterings in the strongly coupled EWSB sector.
For a $~2\rightarrow n-2~$
( $n\geq 4$ ) strong pure $V_L$-scattering process,
the $B$-term defined in (1.1) is  of order$~~O(g^2)v^{n-4}~~$, 
where $~v=246$~GeV. By the electroweak power counting
analysis~\cite{et5,et4}, it has been shown~\cite{et4} that all
$g$-dependent contributions from either vertices or the mass poles of
gauge-boson, Goldstone boson and ghost fields are at most
of $~O(g^2)~$ and the contributions of fermion Yukawa couplings ($y_f$)
coming from fermion-loops are of $~O(\frac{y_f^2}{16\pi^2})\leq
O(\frac{g^2}{16\pi^2})~$ since $~y_f\leq y_t \simeq O(g)~$.
Also, in the factor $C^a_{\rm mod}$ all loop-level $\Delta^a_i$-quantities 
[cf. eq.~(2.9) and Fig.1] are of $~~O(\frac{g^2}{16\pi^2})~~$ 
since they contain at least two ghost-gauge-boson or ghost-scalar
vertices. Hence, if the
$~B$-term (of $~O(g^2)f_\pi^{n-4}~$) is ignored in the strong pure
$V_L$-scattering amplitude, all other $g$- and $y_f$-dependent terms
should also be ignored. Consequently we can simplify
the modification factor such that $~C^a_{\rm mod} \simeq 1+O(g^2)~$ 
by choosing~\cite{et4}
$$
\begin{array}{l}
\displaystyle
\left.
Z_{\phi^a}=\left[\left(\displaystyle\frac{M_V}{M_V^{\rm phys}}\right)^2
Z_{V^a}Z_{M_V}^2 \right]\right|_{g,e,y_f=0} ~~,~~~~(~{\it Scheme-III}~)~.
\end{array}
\eqno(3.22)   
$$
All other renormalization conditions
can be freely chosen as in any standard renormalization scheme.
(Here $~M_V^{\rm phys}~$ is the physical mass pole of the 
gauge boson $V^a$. Note that we have set $~M_V^{\rm phys}=M_V~$ in 
{\it Scheme-IV}~ for simplicity.) In this scheme,
because of the neglect of all gauge and Yukawa couplings,
 all gauge-boson, Goldstone-boson and ghost mass poles are
approximately zero. Thus, all $R_\xi$-gauges (including both
't~Hooft-Feynman and Landau gauges) become {\it equivalent}, for
the case of strong pure $V_L$-scatterings in both the
heavy Higgs SM or the EWCL formalism. But, for 
processes involving fields other than longitudinal gauge bosons,
only {\it Scheme-II}~ and {\it Scheme-IV}~
are suitable.\footnote{Some interesting examples are 
$~W_LW_L,Z_LZ_L\rightarrow t\bar{t}~$, $~V_LH\rightarrow V_LH~$, and 
$~AA\rightarrow W_LW_L, WWV_LV_L~$, etc. ($A=$photon).}~~  
Even in the case of pure $V_L$-scattering, we
note that in the kinematic regions around
the $t$ and $u$ channel singularities, photon exchange
becomes important and must be retained~\cite{et5}.
In this case, {\it Scheme-IV}~ or {\it Scheme-II}~ is required to remove
the $~C^a_{\rm mod}$-factors. 

In summary, renormalization {\it Scheme-IV} ensures the
modification-free formulation of the ET [cf. (3.1)-(3.5)].  It is
valid for all $R_\xi$-gauges, but is particularly convenient for the
Landau gauge where all unphysical Goldstone boson and 
ghost fields are exactly massless~\cite{et2,MW}.
{\it Scheme-II}~\cite{et1,et2}, on the other hand, is
particularly convenient for 't~Hooft-Feynman gauge.
For all other $R_\xi$-gauges,
both schemes are valid,  but the {\it Scheme-IV}~ may be more
convenient due to the absence of the tree-level gauge-Goldstone boson
mixing.

\section{Explicit One-Loop Calculations}

\subsection{One-loop Calculations for the Heavy Higgs Standard Model}

To demonstrate the effectiveness of {\it Scheme-IV}~, we first
consider the 
heavy Higgs limit of the standard model.  The complete one-loop
calculations for proper self-energies and renormalization constants in
the heavy Higgs limit have been given for general $R_\xi$-gauges in
reference~\cite{et2} for renormalization {\it Scheme-I}~. Since, in
this scheme, 
$~\Omega^{W,ZZ}_{\xi,\kappa}=1+\delta\Omega^{W,ZZ}_{\xi,\kappa} 
\approx 1$~ at one-loop in the heavy Higgs limit, {\it Scheme-I}~
coincides with {\it Scheme-IV}~ to this order.  Hence, we can
directly use those results to demonstrate that $C^{W,Z}_{mod}$ is
equal to unity in {\it Scheme-IV}~.  
The results for the charged and neutral sectors are~\cite{et2}:
$$
\begin{array}{rcl}
\displaystyle
     \widetilde{\Pi}_{WW,0}(k^2) & 
= &  \displaystyle  -{g^2\over16\pi^2}\left[{
   1\over8}m_H^2+{3\over4}M_W^2\ln{m_H^2\over M_W^2}\right]~~,\\[0.5cm]
\displaystyle
     \widetilde{\Pi}_{ZZ,0}(k^2) & 
= &    \displaystyle    -{g^2\over{16\pi^2c_{\rm w}^2}}\left[{
   1\over8}m_H^2+{3\over4}M_Z^2\ln{m_H^2\over M_Z^2}\right]~~,\\[0.5cm]
\displaystyle
     \widetilde{\Pi}_{\phi^\pm\phi^\pm,0}(k^2) & 
= &   \displaystyle -{g^2\over16\pi^2}
         k^2\left[{1\over8}{m_H^2\over M_W^2} + \left({3\over4} - 
         {\xi_W\over2}\right)\ln{m_H^2\over M_W^2}\right]~~,\\[0.5cm]
\displaystyle
     \widetilde{\Pi}_{\phi^Z\phi^Z,0}(k^2) & 
= &        \displaystyle -{g^2\over{16\pi^2c_{\rm w}^2}}
         k^2\left[{1\over8}{m_H^2\over M_Z^2} + \left({3\over4} - 
         {\xi_Z\over2}\right)\ln{m_H^2\over M_Z^2}\right]~~,\\[0.5cm]
\displaystyle
     \delta Z_{M_W} & 
= &   \displaystyle  -{g^2\over16\pi^2}\left[{1\over16}
   {m_H^2\over M_W^2}+{5\over12}\ln{m_H^2\over M_W^2}\right]~~,\\[0.5cm]
\displaystyle
     \delta Z_{M_Z} & 
= &   \displaystyle -{g^2\over16\pi^2c_{\rm w}^2}\left[{1\over16}
   {m_H^2\over M_Z^2}+{5\over12}\ln{m_H^2\over M_Z^2}\right]~~,\\[0.5cm]
\displaystyle
     \delta Z_{W} & 
= &   \displaystyle  -{g^2\over16\pi^2}{1\over12}
         \ln{m_H^2\over M_W^2}~~,\\[0.5cm]
\displaystyle 
        \delta Z_{ZZ} & 
= &   \displaystyle  {-}{g^2\over16\pi^2}{1\over12c_{\rm w}^2}
         \ln{m_H^2\over M_Z^2}~~,\\[0.5cm]
\displaystyle
  \delta Z_{\phi^\pm} & = &    
\displaystyle  -{g^2\over16\pi^2}\left[
         {1\over8}{m_H^2\over M_W^2} + \left({3\over4}-{\xi_W\over2}
         \right)\ln{m_H^2\over M_W^2}\right]~~,\\[0.5cm]
\displaystyle
 \delta Z_{\phi^Z} & = &   
\displaystyle  -{g^2\over16\pi^2c_{\rm w}^2}\left[
         {1\over8}{m_H^2\over M_W^2} + \left({3\over4}-{\xi_Z\over2}
         \right)\ln{m_H^2\over M_Z^2}\right]~~,\\[0.5cm]
\displaystyle
     \Delta^{W}_1(k^2) & = &   
\displaystyle {-}{g^2\over16\pi^2}{\xi_W\over4}
         \ln{m_H^2\over M_W^2}~~,\\[0.5cm]
\displaystyle
     \Delta^{ZZ}_1(k^2) & = &   
\displaystyle {-}{g^2\over16\pi^2}{\xi_Z\over4c_{\rm w}^2}
         \ln{m_H^2\over M_Z^2}~~.\\[0.5cm]
\end{array}
\eqno(4.1)     
$$
Note that $\Delta^W_{2,3}$ and the corresponding  neutral sector terms
(cf. Fig.~1) are not enhanced by powers or logarithms of the large
Higgs mass, and thus are ignored in this approximation.  Substituting
$~\delta Z_W$, $\delta Z_{M_W}$, $\widetilde{\Pi}_{\phi^\pm\phi^\pm
,0}$, $\widetilde{\Pi}_{WW ,0}~$ and 
$~\delta Z_{ZZ}$, $\delta Z_{M_Z}$, $\widetilde{\Pi}_{\phi^Z\phi^Z
,0}$, $\widetilde{\Pi}_{ZZ ,0}~$ 
into the right hand sides of
eqs.~(2.41) and (2.42) respectively, we obtain
$$
\begin{array}{ll}
\delta Z_{\phi^\pm} & 
\displaystyle =~\displaystyle{g^2\over16\pi^2}\left[-{1\over8}
        {m_H^2\over M_W^2}+\left(-{3\over4}+{\xi_W\over2}\right)
        \ln{m_H^2\over M_W^2}\right]~~,\\[0.5cm]
\delta Z_{\phi^Z} & =~\displaystyle{g^2\over16\pi^2c^2_{\rm w}}
        \left[-{1\over8}
        {m_H^2\over M_W^2}+\left(-{3\over4}+{\xi_Z\over2}\right)
        \ln{m_H^2\over M_Z^2}\right]~~,
\end{array}
\eqno(4.2)   
$$
verifying the equivalence of {\it Schemes-I}~ and {\it -IV}~ in
this limit.  This means that the one-loop value of ~$C_{\rm
mod}^{W,Z}$~ should be equal to unity.  
Using (2.8), (2.11) and the
renormalization constants given in (4.1), we directly compute the
~$C_{\rm mod}^{W,Z}$~ up to one-loop
in the $R_\xi$-gauges for the heavy Higgs case as
$$
\begin{array}{ll}
\displaystyle
     C_{\rm mod}^W &
\displaystyle = ~1+{1\over2}\left(\delta Z_W -
        \delta Z_{\phi^\pm} +2\delta Z_{M_W}\right)
      + \Delta^{W}_{1}(M_W^2)\\[0.5cm]
\displaystyle
     & 
\displaystyle =~1 + {g^2\over16\pi^2}\left\{\left({1\over16} - {1\over16}
         \right){m_H^2\over M_W^2} + \left({1\over24} + {3\over8} -
        {5\over12} - {\xi_W\over4}+ {\xi_W\over4}\right)
        \ln{m_H^2\over M_W^2}\right\}\\[0.5cm]
     & =~ 1 + O(2~{\rm loop})~~,\\[0.5cm]
\displaystyle
     C_{\rm mod}^Z &
\displaystyle = ~1+{1\over2}\left(\delta Z_{ZZ} -
        \delta Z_{\phi^Z} +2\delta Z_{M_Z}\right)
      + \Delta^{Z}_{1}(M_Z^2)\\[0.5cm]
\displaystyle
     & \displaystyle 
=~1 + {g^2\over16\pi^2c_{\rm w}^2}\left\{\left({1\over16} - {1\over16}
         \right){m_H^2\over M_W^2} + \left({1\over24} + {3\over8} -
        {5\over12} - {\xi_Z\over4}+ {\xi_Z\over4}\right)
        \ln{m_H^2\over M_Z^2}\right\}\\[0.5cm]
     & =~ 1 + O(2~{\rm loop})~~. \\[0.5cm]
\end{array}
\eqno(4.3)   
$$
Equation (4.3) is an explicit one-loop proof that $~C^{W,Z}_{mod}=1~$ in
{\it Scheme-IV}~.  The agreement of {\it Schemes-I}~ and {\it -IV}~
only occurs in the heavy Higgs limit up to one-loop order.  
When the Higgs is not very heavy,
the full one-loop corrections from all scalar and gauge couplings must
be included, so that {\it Scheme-IV}~ and {\it Scheme-I}~ are
no longer equivalent.

\subsection{Complete One-Loop Calculations for the $U(1)$ Higgs theory}

The simplest case to explicitly demonstrate {\it Scheme-IV}~ for
arbitrary Higgs mass is the $U(1)$ Higgs theory.  In this section, we
use complete one-loop calculations in the $U(1)$ Higgs theory (for any
value of $m_H$) to explicitly verify that $C_{mod}=1$ in {\it
Scheme-IV}~ for both Landau and 't~Hooft-Feynman gauges.

The $U(1)$ Higgs theory contains minimal field content: 
the physical Abelian gauge field $A_\mu$ (with mass $M$), the Higgs
field $H$ (with mass $m_H$), 
as well as the unphysical Goldstone boson field
$\phi$ and the Faddeev-Popov ghost fields $c, \overline{c}$
(with mass poles at $\xi\kappa M$).  Because the symmetry group is
Abelian, $\Delta_2$ and $\Delta_3$ do not occur and the modification
factor is given by
$$
	C_{\rm mod} = \left({Z_A\over Z_\phi}\right)^{1\over2}
	Z_M[1+\Delta_1(M^2)]~~,
\eqno(4.4)
$$
with
$$
\Delta_1(k^2)=\displaystyle\frac{Z_gZ_H^{\frac{1}{2}}}{Z_M}
\frac{g\mu^{\epsilon}}{M}
\int_q <0|H(-k-q)c(q)|\bar{c}(k)>
\eqno(4.5)           
$$
where $~~\int_q \equiv \displaystyle\int\frac{dq^D}{(2\pi)^D}~~$
and $~D=4-2\epsilon~$. $~\Delta_1~$ vanishes identically in Landau
gauge ($~\xi =0~$), because in the $U(1)$ theory the ghosts
couple only to the Higgs boson and that coupling is proportional to
$~\xi~$. In {\it Scheme-IV}~, the wavefunction renormalization
constant $Z_\phi$ of the Goldstone boson field is defined to be
$$
Z_\phi \displaystyle 
= Z_A{Z_{M}^2M^2-\widetilde{\Pi}_{AA,0}(M^2)\over
	M^2-\widetilde{\Pi}_{\phi\phi,0}(M^2)}~~.
\eqno(4.6)
$$
Substituting (4.6) into (4.5), we obtain the following one-loop
expression for $C_{\rm mod}$
$$
\displaystyle
C_{\rm mod} \displaystyle 
= 1 + {1\over2}M^{-2}\left[\widetilde{\Pi}_{AA,0}(M^2)-
                \widetilde{\Pi}_{\phi\phi,0}(M^2)\right]+\Delta_1(M^2)~~.
\eqno(4.7)
$$
We shall now explicitly verify that $C_{mod}$ is equal to unity in
both Landau and 't~Hooft-Feynman gauges.

In Landau gauge:\\[0.35cm]
$$
\begin{array}{rl}
\displaystyle
     \left.\widetilde{\Pi}_{AA,0}(k^2)\right|_{\xi=0}
= ig^2 &
\displaystyle \left\{-I_1(m_H^2)
           \vphantom{\left\lgroup I_{41}(k^2;M^2,m_H^2)\right\rgroup}
     -4M^2I_2(k^2;M^2,m_H^2) + k^2I_2(k^2;0,m_H^2)\right. \\[0.5cm]
\displaystyle
     &\left. +4k^2I_3(k^2;0,m_H^2) + 4\left\lgroup
          I_{41}(k^2;M^2,m_H^2) + k^2I_{42}(k^2;a^2,b^2)
          \right\rgroup\right\}  ~~,\\[0.5cm]
\displaystyle
     \left.\widetilde{\Pi}_{\phi\phi,0}(k^2)\right|_{\xi=0}
          = ig^2
&\displaystyle \left\{
          -{m_H^2\over M^2}I_1(m_H^2) + {m_H^4\over M^2}
          I_2(k^2;0,m_H^2) - 4k^2I_2(k^2;M^2,m_H^2)\right. \\[0.5cm]
\displaystyle
     &\displaystyle \left.\kern60pt + 4{k^2\over M^2}\left\lgroup I_{41}
          (k^2;M^2,m_H^2) - I_{41}(k^2;0,m_H^2)\right\rgroup\right.  
   \\[0.5cm]
     &
\displaystyle 
\left.\kern60pt + 4{m_H^4\over M^2}\left\lgroup I_{42}(k^2;M^2,
          m_H^2) - I_{42}(k^2,0,m_H^2)\right\rgroup\right\}~~,\\[0.5cm]
\displaystyle
     \left.\Delta_1(k^2)\vphantom{\widetilde{\Pi}_{\phi\phi,0}}
          \right|_{\xi=0}\kern-5pt = 0  ~~, & \\[0.35cm]
\end{array}
\eqno(4.8)
$$
where the quantities $I_j$'s denote the one-loop integrals:
$$
\begin{array}{rl}
\displaystyle
    I_1(a^2) &
\displaystyle =~\mu^{2\epsilon}\int{d^{4-2\epsilon}p
        \over(2\pi)^{4-2\epsilon}}{1\over p^2-a}
        \equiv\mu^{2\epsilon}\int_p{1\over p^2-a} ~~,\\[0.5cm]
\displaystyle 
   I_2(k^2;a^2,b^2) &
\displaystyle =~ \mu^{2\epsilon}\int_p
                 {1\over(p^2-a)[(p+k)^2-b]} ~~,\\[0.5cm]
\displaystyle
    I_3^{\mu}(k;a^2,b^2) &
\displaystyle =~ \mu^{2\epsilon}\int_p{p^\mu\over
         (p^2-a)[(p+k)^2-b]}= k^\mu I_3(k^2;a^2,b^2) ~~,\\[0.5cm]
\displaystyle 
   I_4^{\mu\nu}(k;a^2,b^2) &
\displaystyle =~ \mu^{2\epsilon}\int_p
         {p^\mu p^\nu\over(p^2-a)[(p+k)^2-b]}
         = g^{\mu\nu} I_{41}(k^2;a^2,b^2) + 
         k^\mu k^\nu I_{42}(k^2;a^2,b^2) ~~,\\[0.35cm]
\end{array}
\eqno(4.9)   
$$
which are evaluated in Appendix-C.  Substituting (4.9) into the
right hand side of (4.7), we obtain 
$$
C_{\rm mod}=1+~O(2~{\rm loop})~~,~~~~~(~{\rm in~Landau~gauge}~)~~,
\eqno(4.10)   
$$
as expected.

We next consider the 't~Hooft-Feynman gauge in which
$~\Delta_1(M^2)~$ is non-vanishing:
$$
\begin{array}{rl}
\displaystyle
    \left.\widetilde{\Pi}_{AA,0}(k^2)\right|_{\xi=1}
           &
\displaystyle =~ig^2\left\{-I_1(m_H^2)-I_1(M^2)
          +(k^2-4M^2)I_2 +4k^2I_3+4(I_{41}+k^2I_{42})\right\}~,\\[0.5cm]
\displaystyle 
   \left.\widetilde{\Pi}_{\phi\phi,0}(k^2)\right|_{\xi=1}
            &
\displaystyle =~ig^2\left\{\left(1+{m_H^2\over M^2}\right)\left\lgroup
            I_1(M^2)-I_1(m_H^2)\right\rgroup + \left({m_H^4\over M^2}
            -M^2-4k^2\right)I_2 -4k^2I_3 \right\} ~,\\[0.5cm]
\displaystyle
    \left.\Delta_1(k^2)\vphantom{\widetilde{\Pi}_{\phi\phi,0}}
          \right|_{\xi=1}
           &
\displaystyle =~ig^2I_2(M^2;M^2,m_H^2) ~,\\[0.3cm]
\end{array}
\eqno(4.11)
$$
where $~I_j = I_j(k^2;M^2,m_H^2)~$ for $~j\geq 2~$.
Again, we substitute (4.11) into equation~(4.7) and find that
$$
C_{\rm mod}=1+~O(2~{\rm loop})~~,~~~~~(~{\rm in~'t~Hooft-Feynman~gauge}~)~~.
\eqno(4.12)   
$$

\section{Conclusions}

In this paper, we have constructed a convenient new renormalization
scheme, called {\it Scheme-IV}, which rigorously reduces all radiative
modification factors to the equivalence theorem ($~C^a_{\rm mod}$~'s)
to unity in all $R_\xi$-gauges including both 't~Hooft-Feynman and
Landau gauges.  
This new {\it Scheme-IV}~ proves particularly convenient for
Landau gauge which cannot be included in the previously described {\it
Scheme-II}~\cite{et1,et2}. Our formulation is explicitly constructed
for both the $SU(2)_L$ and $SU(2)_L\otimes U(1)_Y$ theories
[cf. sections~2 and 3]. Furthermore, we have generalized our formulation
to the important effective Lagrangian formalisms for both
the non-linear~\cite{app} and linear~\cite{linear} realizations of the
electroweak symmetry breaking (EWSB) sector, where the new physics
(due to either strongly or weakly coupled EWSB mechanisms)
has been parameterized by effective operators (cf. section~3.2 and
Appendix-A).
In the construction of the {\it Scheme-IV}~ 
(cf. section~2.2), we first re-express
the $~C^a_{\rm mod}$-factors in terms of proper self-energies of the
unphysical sector by means of the $R_\xi$-gauge WT identities. Then,
we simplify the $~C^a_{\rm mod}$-factors to unity by
specifying the subtraction condition for the Goldstone boson wavefunction 
renormalization constant $Z_{\phi^a}$ [cf. (2.28) and (2.41-42)].
This choice for $Z_{\phi^a}$ 
is determined by the known gauge and Goldstone boson self-energies
(plus the gauge boson wavefunction and mass renormalization constants)
which must be computed in any practical renormalization scheme.
We emphasize that the implementation of the {\it Scheme-IV}~
requires no additional calculation (of $\Delta^a_i$-quantities, for
instance) beyond what is required for the {\it standard} 
radiative correction computations~\cite{Hollik}.  
Based upon this radiative modification-free formulation for the
equivalence theorem [cf. (3.4)], we have further proposed
a new prescription, which we call the `` Divided Equivalence Theorem '' 
(DET) [cf. (3.14)-(3.15) and discussions followed], 
for minimizing the approximation due to ignoring the additive $B$-term
in the equivalence theorem (3.4) or (1.1).
Finally, we have explicitly verified that, 
in {\it Scheme-IV}, the $~C^a_{\rm mod}$-factor is 
reduced to unity in the heavy Higgs limit of the standard model 
(cf. section~4.1) and for arbitrary Higgs mass in the $U(1)$ Higgs theory
(cf. section~4.2).  

\Ack
We thank Michael Chanowitz, Yu-Ping Kuang, C.--P. Yuan and Peter Zerwas for 
carefully reading the manuscript, and for their useful suggestions and
support.  H.J.H is supported by the AvH of Germany.  \DOE

\appendix{A}{Next-to-leading Order Effective Operators in the EWCL}

Within the EWCL formalism, the next-to-leading order 
effective operators arising from new physics can be 
parameterized as~\cite{app,et5} 
$$
\begin{array}{ll}
{\cal L}_{\rm eff}^{\prime} &\displaystyle  
\equiv ~{\cal L}_{\rm GB}^{\prime}+{\cal L}_{\rm F}^{\prime}
\end{array}
\eqno(A1)
$$
The bosonic part $~{\cal L}_{\rm GB}^{\prime}~$ is given by
$$
\begin{array}{lcl}
 {\cal L}_{\rm GB}^{\prime}~ & = &
\displaystyle  {\cal L}^{(2)\prime}+
             \displaystyle\sum_{n=1}^{14} {\cal L}_n ~~,\\[0.4cm]
{\cal L}^{(2)\prime} & = & \ell_0 (\frac{v}{\Lambda})^2~\frac{v^2}{4}
               [ {\rm Tr}({\cal T}{\cal V}_{\mu})]^2 ~~,\\[0.2cm]
{\cal L}_1 & = & \ell_1 (\frac{v}{\Lambda})^2~ \frac{gg^\prime}{2}
B_{\mu\nu} {\rm Tr}({\cal T}{\bf W^{\mu\nu}}) ~~,\\[0.2cm]
{\cal L}_2 & = & \ell_2 (\frac{v}{\Lambda})^2 ~\frac{ig^{\prime}}{2}
B_{\mu\nu} {\rm Tr}({\cal T}[{\cal V}^\mu,{\cal V}^\nu ]) ~~,\\[0.2cm]
{\cal L}_3 & = & \ell_3 (\frac{v}{\Lambda})^2 ~ig
{\rm Tr}({\bf W}_{\mu\nu}[{\cal V}^\mu,{\cal V}^{\nu} ]) ~~,\\[0.2cm]
{\cal L}_4 & = & \ell_4 (\frac{v}{\Lambda})^2 
                     [{\rm Tr}({\cal V}_{\mu}{\cal V}_\nu )]^2 ~~,\\ [0.2cm] 
{\cal L}_5 & = & \ell_5 (\frac{v}{\Lambda})^2 
                     [{\rm Tr}({\cal V}_{\mu}{\cal V}^\mu )]^2 ~~,\\  [0.2cm]
{\cal L}_6 & = & \ell_6 (\frac{v}{\Lambda})^2 
[{\rm Tr}({\cal V}_{\mu}{\cal V}_\nu )]
{\rm Tr}({\cal T}{\cal V}^\mu){\rm Tr}({\cal T}{\cal V}^\nu) ~~,\\[0.2cm]
{\cal L}_7 & = & \ell_7 (\frac{v}{\Lambda})^2 
[{\rm Tr}({\cal V}_\mu{\cal V}^\mu )]
{\rm Tr}({\cal T}{\cal V}_\nu){\rm Tr}({\cal T}{\cal V}^\nu) ~~,\\[0.2cm]
{\cal L}_8 & = & \ell_8 (\frac{v}{\Lambda})^2~\frac{g^2}{4} 
[{\rm Tr}({\cal T}{\bf W}_{\mu\nu} )]^2  ~~,\\[0.2cm]
{\cal L}_9 & = & \ell_9 (\frac{v}{\Lambda})^2 ~\frac{ig}{2}
{\rm Tr}({\cal T}{\bf W}_{\mu\nu}){\rm Tr}
        ({\cal T}[{\cal V}^\mu,{\cal V}^\nu ]) ~~,\\[0.2cm]
{\cal L}_{10} & = & \ell_{10} (\frac{v}{\Lambda})^2\frac{1}{2}
[{\rm Tr}({\cal T}{\cal V}^\mu){\rm Tr}({\cal T}{\cal V}^{\nu})]^2 ~~,\\[0.2cm]
{\cal L}_{11} & = & \ell_{11} (\frac{v}{\Lambda})^2 
~g\epsilon^{\mu\nu\rho\lambda}
{\rm Tr}({\cal T}{\cal V}_{\mu}){\rm Tr}
({\cal V}_\nu {\bf W}_{\rho\lambda}) ~~,\\[0.2cm]
{\cal L}_{12} & = & \ell_{12}(\frac{v}{\Lambda})^2 ~2g
                    {\rm Tr}({\cal T}{\cal V}_{\mu}){\rm Tr}
                  ({\cal V}_\nu {\bf W}^{\mu\nu}) ~~,\\[0.2cm]
{\cal L}_{13} & = & \ell_{13}(\frac{v}{\Lambda})^2~ 
      \frac{gg^\prime}{4}\epsilon^{\mu\nu\rho\lambda}
      B_{\mu\nu} {\rm Tr}({\cal T}{\bf W}_{\rho\lambda}) ~~,\\[0.2cm]     
{\cal L}_{14} & = & \ell_{14} (\frac{v}{\Lambda})^2~\frac{g^2}{8} 
\epsilon^{\mu\nu\rho\lambda}{\rm Tr}({\cal T}{\bf W}_{\mu\nu})
{\rm Tr}({\cal T}{\bf W}_{\rho\lambda})   ~~,\\[0.2cm]
\end{array}
\eqno(A2)                                         
$$
with
$$
{\cal V}_{\mu}\equiv (D_{\mu}U)U^\dagger~~,~~~~
{\cal T}\equiv U\tau_3 U^{\dagger} ~~,
\eqno(A3)                                         
$$
where $~{\cal T}~$ is the custodial $SU(2)_C$-violating operator.
In $(A2)$, $~{\cal L}^{(2)\prime}~$ and  
$~{\cal L}_1\sim {\cal L}_{11}~$ are $CP$-conserving while 
$~{\cal L}_{12}\sim {\cal L}_{14}~$ are
$CP$-violating. Many of these effective operators can 
be probed at the LHC and LC 
via longitudinal gauge boson  scattering processes~\cite{et5,LHC,LC}.
For the fermionic part $~{\cal L}_{\rm F}^{\prime}~$, we refer the reader
to the reference~\cite{fermion} for details.

\appendix{B}{Non-linear BRST Transformations of
Goldstone Boson Fields and the Faddeev-Popov Term in the EWCL}

In this Appendix, we first summarize the derivation for the
non-linear $SU(2)_L\otimes U(1)_Y$ BRST transformations of the
Goldstone boson fields [cf. (3.8)] and then give the complete
Landau gauge Faddeev-Popov term.

For simplicity of notation, we need only derive
the usual $SU(2)_L\otimes U(1)_Y$ gauge transformations for
the $~\pi^a$~'s, since their BRST transformations
are obtained by simply replacing the usual gauge  parameters
($\theta^a(x)$) by the corresponding ghost fields
$~c^a(x)~$.
Consider the infinitesimal $SU(2)_L\otimes U(1)_Y$ gauge transformation
for the $U$-matrix [cf. (3.7)]:
$$
\begin{array}{ll}
U~\Longrightarrow ~U^{\prime} 
& \displaystyle =~{\cal G}_L(\theta_L) U {\cal G}_Y^{\dagger}(\theta_Y)
\equiv U+\delta U\\[0.35cm]
& \displaystyle =~U+\frac{ig}{2}\theta_L^a\tau^aU-\frac{ig'}{2}\theta_YU\tau^3
+~O(\theta_{L}^2,~\theta_{Y}^2)
\end{array}
\eqno(B1)   
$$
with 
$$
\begin{array}{ll}
{\cal G}_L & 
\displaystyle
=~\exp \left[ig\theta^a_L(x)\tau^a/2\right]
~=~1+ig\theta^a_L\tau^a/2 + ~O(\theta_L^2)~\in~ SU(2)_L ~~,\\[0.3cm]
{\cal G}_Y & 
\displaystyle
=~\exp \left[ig\theta_Y(x)\tau^3/2\right]
~=~1+ig^{\prime}\theta_Y\tau^3/2 + ~O(\theta_Y^2)~\in~ U(1)_Y ~~.
\end{array}
\eqno(B2)
$$
Expanding the $U$-matrix in $(B1)$, we obtain
$$
\begin{array}{l}
\displaystyle
\delta U ~=~\\[0.5cm]
\displaystyle
{g\over2}\cos\underline{\pi}\left[ \begin{array}{ll}
\theta_L^3\left(i-\omega\underline{\pi}_3\right)+
\left(\theta_L^1-i\theta_L^2\right)(-\underline{\pi}_1-i\underline{\pi}_2)
\omega
~&~  
\displaystyle
\theta_L^3\left(-\underline{\pi}_1+i\underline{\pi}_2\right)\omega +
\left(\theta_L^1-i\theta_L^2\right)
\left(i+\omega\underline{\pi}_3\right)\\[0.5cm]
\displaystyle
\theta_L^3\left(\underline{\pi}_1+i\underline{\pi}_2\right)\omega +
\left(\theta_L^1+i\theta_L^2)(i-\omega\underline{\pi}_3\right)
~&~
\displaystyle
\theta_L^3\left(-i-\omega\underline{\pi}_3\right)+
\left(\theta_L^1+i\theta_L^2\right)(-\underline{\pi}_1+i\underline{\pi}_2)
\omega
\end{array} \right]\\[0.9cm]
\displaystyle
~+~ {g'\over2}\cos\underline{\pi}\left[ \begin{array}{ll}
(-i+\omega\underline{\pi}_3)\theta_Y   
~~&~~
\displaystyle
\omega (-\underline{\pi}_1+i\underline{\pi}_2)\theta_Y\\[0.5cm]
\omega (\underline{\pi}_1+i\underline{\pi}_2)\theta_Y
~~&~~
\displaystyle
(i+\omega\underline{\pi}_3)\theta_Y \end{array} \right]
\end{array}
\eqno(B3)    
$$
where $~~\omega \equiv \zeta^{-1}~~$ and 
all notations (including $~\zeta~$) are defined in (3.9).

By differential variation of ~$U$~ with respect to the $~\pi^a$-field,
we obtain 
$$
\delta U~=~\displaystyle\frac{\sin\underline{\pi}}{\underline{\pi}}
\left[ \begin{array}{ll}
\displaystyle
-\vec{\underline{\pi}}\cdot\delta\vec{\underline{\pi}}
+i\left(\eta \underline{\pi}_3\vec{\underline{\pi}}
\cdot\delta\vec{\underline{\pi}}+\delta\underline{\pi}_3\right) ~~&~~
\eta\left(i\underline{\pi}_1+\underline{\pi}_2\right)
\vec{\underline{\pi}}\cdot\delta\vec{\underline{\pi}}+
\left(i\delta\underline{\pi}_1+\delta\underline{\pi}_2\right)\\[0.5cm]
\displaystyle
\eta\left(i\underline{\pi}_1-\underline{\pi}_2\right)
\vec{\underline{\pi}}\cdot\delta\vec{\underline{\pi}}+
\left(i\delta\underline{\pi}_1-\delta\underline{\pi}_2\right)
~~&~~ -\vec{\underline{\pi}}\cdot\delta\vec{\underline{\pi}}
-i\left(\eta \underline{\pi}_3\vec{\underline{\pi}}
\cdot\delta\vec{\underline{\pi}}+\delta\underline{\pi}_3\right)
\end{array}  \right]
\eqno(B4)    
$$
where 
$$
\displaystyle
\pi^\pm =\frac{1}{\sqrt{2}}\left(\pi_1\mp i\pi_2\right)~,~~~~
\pi^Z = \pi_3~.
\eqno(B5)    
$$

Equating the two expressions for $~\delta U~$ in $(B3)$ and $(B4)$, we 
 derive the BRST transformations for $~\pi^\pm~$ and $~\pi^Z~$
given in (3.8). The BRST transformations for the
gauge fields [cf. (3.8)] are the same as in the SM.

After setting up the BRST transformations (3.8), we can 
straightforwardly write down the complete $R_\xi$-gauge 
Faddeev-Popov term for the EWCL from (3.10).
Since {\it Scheme-IV}~ is particularly useful for Landau gauge, 
we give only the Landau gauge Faddeev-Popov term. From (3.10) and
(3.8), we obtain, in Landau gauge,
$$
\begin{array}{ll}
{\cal L}_{\rm FP} \displaystyle
~= & -\bar{c}^+\partial^2 c^++\bar{c}^+i\partial^\mu\left[
e\left(W^+_\mu c^A-A_\mu c^+\right)+g{\rm c}_{\rm w}
\left(W^+_\mu c^Z-Z_\mu c^+\right)\right]\\[0.3cm]
& -\bar{c}^-\partial^2 c^--\bar{c}^-i\partial^\mu\left[
e\left( W^-_\mu c^A-A_\mu c^-\right)+g{\rm c}_{\rm w}
\left(W^-_\mu c^Z-Z_\mu c^-\right)\right]\\[0.3cm]
& -\bar{c}^Z\partial^2 c^Z+ig{\rm c}_{\rm w}\bar{c}^Z\partial^\mu
\left[ W^-_\mu c^+-W^+_\mu c^-\right]\\[0.3cm]
& -\bar{c}^A\partial^2 c^A + ie\bar{c}^A\partial^\mu
\left[ W^-_\mu c^+-W^+_\mu c^-\right]   ~~~.
\end{array}
\eqno(B6)   
$$
Finally, we remark that, 
although the Faddeev-Popov term in 't~Hooft-Feynman gauge 
is much more complicated than  in Landau gauge,
it is still
useful due to the  simplicity of the gauge boson propagators.
The main advantage of the new {\it Scheme-IV}~ is 
its applicability to {\it all} $R_\xi$-gauges including both 
Landau and 't~Hooft-Feynman gauges. 

\appendix{C}{Analytic Expressions for the One-Loop Integrals}

Finally, we give the complete analytic expressions for the one-loop
integrals ($I_i$~'s) used in section~4.2 [cf. (4.9)] 
for explicit calculations. For simplicity of notation, we define 
$$
\displaystyle
\frac{1}{\hat{\epsilon}}~
\equiv~\frac{1}{\epsilon} -\gamma +\ln (4\pi \mu^2) ~~.
\eqno(C1)   
$$
Then, we have
$$
\begin{array}{ll}
I_1(a^2) & =~\displaystyle\mu^{2\epsilon}\int_p{1\over p^2-a}~
 =~\displaystyle\frac{i}{16\pi^2}a^2\left[\frac{1}{\hat{\epsilon}}
-\ln a^2 +1\right]  ~~.
\end{array}
\eqno(C2)   
$$
$$
\begin{array}{ll}
\displaystyle 
   I_2(k^2;a^2,b^2) &
\displaystyle =~ \mu^{2\epsilon}\int_p
                 {1\over(p^2-a)[(p+k)^2-b]}\\[0.5cm]
&=~\displaystyle\frac{i}{16\pi^2}\left[\frac{1}{\hat{\epsilon}}
-\ln (ab) +2+\frac{a^2-b^2}{k^2}\ln\frac{b}{a}-\bar{I}_{20}(k^2;a^2,b^2)
\right]  ~~,\\[0.7cm]
\bar{I}_{20}(k^2;a^2,b^2)& \displaystyle =~
   \left\{ \begin{array}{ll}
         -\displaystyle\frac{\sqrt{AB}}{k^2}\ln\frac{\sqrt{-A}+\sqrt{-B}}
  {\sqrt{-A}-\sqrt{-B}}~~,~~~~& (~k^2\leq (a-b)^2~)~~,\\[0.5cm]
  \displaystyle\frac{2\sqrt{-AB}}{k^2}\arctan\sqrt{\frac{B}{-A}}~~,~~~~& 
    (~(a-b)^2<k^2< (a+b)^2~)~~,\\[0.5cm]
 +\displaystyle\frac{\sqrt{AB}}{k^2}\left[\ln\frac{\sqrt{B}+\sqrt{A}}
  {\sqrt{B}-\sqrt{A}}-i\pi\right]~~,~~~~&
    (~k^2\geq (a+b)^2~)~~,\\[0.5cm]
    \end{array} \right.
\end{array}
\eqno(C3)   
$$
where $~~A=k^2-(a+b)^2~~$ and $~~B=k^2-(a-b)^2~~$.
$$
\begin{array}{ll}
\displaystyle 
    I_3^{\mu}(k;a^2,b^2) &
\displaystyle =~ \mu^{2\epsilon}\int_p{p^\mu\over
         (p^2-a)[(p+k)^2-b]} ~=~ k^\mu I_3(k^2;a^2,b^2)  ~~,\\[0.8cm]
I_3(k^2;a^2,b^2)& \displaystyle =~
-\frac{i}{16\pi^2}{a^2\over2}\left[\frac{1}{\hat{\epsilon}}
    -\bar{I}_{30}(k^2;a^2,b^2)\right] ~~,\\[0.6cm]
\displaystyle
\bar{I}_{30}(k^2;a^2,b^2) & =~\displaystyle\ln a^2-2+\frac{b^2-a^2}{k^2}+ 
{1\over2}\left[1+{2b^2\over k^2}-{(a^2-b^2)^2\over k^4}\right]
\ln{b^2\over a^2} \\[0.6cm]
 &\displaystyle ~~~~~+ \left[ 1+ {a^2-b^2\over k^2}\right]
                    \bar{I}_{20}(k^2;a^2,b^2)~~.
\end{array}
\eqno(C4)   
$$
$$
\begin{array}{l}
\displaystyle 
   I_4^{\mu\nu}(k;a^2,b^2) ~
\displaystyle =~ \mu^{2\epsilon}\int_p
         {p^\mu p^\nu\over(p^2-a)[(p+k)^2-b]}
        ~ =~ g^{\mu\nu} I_{41}(k^2;a^2,b^2) + 
         k^\mu k^\nu I_{42}(k^2;a^2,b^2) ~~,\\[0.8cm]
I_{41}(k^2;a^2,b^2)\displaystyle ~=~
\frac{i}{16\pi^2}{1\over4}\left[\left(\frac{1}{\hat{\epsilon}}+
{1\over2}\right)\left(a^2+b^2-{k^2\over 3}\right)-{13\over18}k^2+{11\over6}
(a^2+b^2)-{(a^2-b^2)^2\over 3k^2}{k^2}\right.\\[0.6cm]
\left.\displaystyle
+\left({k^2\over3}-(a^2+b^2)+{a^4-b^4\over k^2}-{(a^2-b^2)^3\over 3k^4}
\right)\ln{b\over a}-\left(a^2+b^2-{k^2\over3}\right)\ln a^2 +{AB\over 3k^2}
\bar{I}_{20}(k^2;a^2,b^2)\right]~,\\[0.6cm]
I_{42}(k^2;a^2,b^2)\displaystyle ~=~
\frac{i}{16\pi^2}{1\over3}\left[\frac{1}{\hat{\epsilon}}+{13\over6}
+{a^2-5b^2\over 2k^2}+{(a^2-b^2)^2\over k^4}-\left(
1+3{b^2\over k^2}+3{b^2(a^2-b^2)\over k^4}\right.\right.\\[0.6cm]
\displaystyle
\left.\left.-{(a^2-b^2)^3\over k^6}\right)\ln\frac{b}{a}
-\ln a^2-\left(1+\frac{a^2-2b^2}{k^2}+{(a^2-b^2)^2\over k^4}\right)
\bar{I}_{20}(k^2;a^2,b^2)\right] ~~.\\[0.6cm]
\end{array}
\eqno(C5)   
$$


\newpage

\end{document}